\documentclass{emulateapj}

\usepackage{graphicx}
\usepackage[hypertex]{hyperref}

\def \nh {N${\rm _H}$}
 
\def \arcmin {\hbox{$^\prime$}} 
\def \arcsec {\hbox{$^{\prime\prime}$}} 
 
\def\spose#1{\hbox to 0pt{#1\hss}} 
\def\ltsim{$\mathrel{\spose{\lower 3pt\hbox{$\sim$}} 
        \raise 2.0pt\hbox{$<$}}$\thinspace} 
\def\gtsim{$\mathrel{\spose{\lower 3pt\hbox{$\sim$}} 
        \raise 2.0pt\hbox{$>$}}$\thinspace} 
\def \msun {${\rm M_\odot}$} 
 
\def \nh {$N_{\rm H}$}

\def \eg {e.g.}

\def \ie {i.e.} 
 
\def \dtwentyfive {${\rm D_{25}}$}

\newcommand{\pgcx}{${\rm p_{GC-X}}$}

\newcommand{\gcpaper}{Paper~II}

\newcommand{\probhnought}{${\rm prob(H_0)}$}
\newcommand{\zsun}{${\rm Z_\odot}$}
\newcommand{\Sn}{${\rm S_N}$}
\newcommand{\Sl}{${\rm S_L}$}
\newcommand{\Sxn}{${\rm S_{N,X}}$}

\newcommand{\chandra }{{\em Chandra}} 
\newcommand{\hst}{{\em HST}}
\newcommand{\xspec }{{\em Xspec}} 
\newcommand{\acis }{{\em ACIS}} 
\newcommand{\ciao }{{\em CIAO}} 
\newcommand{\caldb }{{\em Caldb}} 
\newcommand{\heasoft }{{\em Heasoft}}

\newcommand{\ergps}{${\rm erg\ s^{-1}}$} 
 
\newcommand{\asca }{{\em ASCA}}

\newcommand{\lx }{${\rm L_X}$}

\newcommand{\lgc}{${\rm L_{GC}}$}

\newcommand{\lv }{${\rm L_V}$}
\newcommand{\lk }{${\rm L_K}$}

\newcommand{\twomass}{{\em 2MASS}}

\newcommand{\lsun }{${\rm L_\odot}$}
 
\newcommand{\ned}{{\em{NED}}} 

\newcommand{\thin}{\thinspace}
\slugcomment{Submitted to the Astrophysical Journal}
\shorttitle{LMXBs in Early-Type galaxies}
\shortauthors{Humphrey \& Buote}
 
\begin{document} 
 
\title{Low-Mass X-ray Binaries and Globular Clusters in Early-Type Galaxies. I. Chandra Observations}
\author {Philip J. Humphrey\altaffilmark{1} and David A. Buote\altaffilmark{1}}
\altaffiltext{1}{Department of Physics and Astronomy, University of California at Irvine, 4129 Frederick Reines Hall, Irvine, CA 92697-4575}
\begin{abstract}
We present a \chandra\ survey of LMXBs in 24 early-type galaxies.
Correcting for detection incompleteness, the X-ray luminosity 
function (XLF) of each galaxy is consistent with a powerlaw
with negative logarithmic differential slope, 
$\beta\sim 2.0$.  However, $\beta$ strongly 
correlates with incompleteness, indicating the XLF flattens at 
low-\lx. The composite XLF is well-fitted by a powerlaw with a break at 
$(2.21^{+0.65}_{-0.56})\times 10^{38}$\ergps\ and $\beta=$1.40$^{+0.10}_{-0.13}$ 
and $2.84^{+0.39}_{-0.30}$ below and above it, respectively.
The break is close to the Eddington limit for a 1.4\msun\
neutron-star, but the XLF shape rules out its
representing the division between neutron-star and black-hole systems. 
Although the XLFs are similar, we find evidence of some
variation between galaxies. The high-\lx\ XLF slope does not correlate
with age, but may correlate with [$\alpha$/Fe].
Considering only LMXBs with \lx$>10^{37}$ \ergps,
matching the LMXBs with globular clusters (GCs)
identified in \hst\ observations of 19 of the galaxies, we find
the probability a GC hosts an LMXB is proportional to 
${L_{GC}^\alpha Z_{Fe}^\gamma}$ where 
$\alpha=1.01\pm0.19$ and $\gamma=0.33\pm0.11$.
Correcting for GC luminosity and colour effects, and detection incompleteness,
we find no evidence that the fraction of LMXBs with 
\lx$> 10^{37}$ \ergps\ in GCs ($40$\%), or 
the fraction of GCs hosting LMXBs ($\sim$6.5\%) varies between galaxies. 
The spatial distribution of LMXBs resembles that of 
GCs, and the specific frequency of LMXBs is proportional to the 
GC specific luminosity, consistent with the hypothesis that all
LMXBs form in GCs. If the LMXB lifetime is $\tau_L$ and the duty cycle
is $F_d$, 
our results imply $\sim 1.5 (\tau_L/10^8 yr)^{-1} F_d^{-1}$ 
LMXBs are formed 
Gyr$^{-1}$ per GC and we place an upper limit of 1 active LMXB
in the field per 3.4$\times 10^{9}$ \lsun\ of V-band luminosity.
\end{abstract}

\keywords{Xrays: galaxies--- galaxies: elliptical and lenticular, cD--- 
Xrays: binaries--- galaxies: star clusters}

\section{Introduction}
As the end-points of stellar evolution, studies of 
X-ray binary (XRB) populations
provide a valuable insight not only into black-hole and
neutron-star demographics but also into the history of star-formation
within galaxies. In the four decades since the discovery 
of Sco\thin X-1 \citep{giacconi62a}, generations of 
X-ray satellites have revealed a rich variety of phenomenology
in the $\sim$300 XRBs in the Milky Way \citep{white95,liu00,liu01}.
Unfortunately, studies of these objects are hampered by a number
of factors such as uncertainty in the distances to (and hence
luminosities of) Galactic sources, the limited numbers of
sources and the mixture of old  and young stellar populations in the 
Milky Way, which can make  isolating different influences
on the population challenging.
These problems can, in part, be mitigated through observations of 
external, early-type galaxies, which comprise a clean
old stellar population and XRBs all at essentially the same
distance. The absence of massive, young stars in these systems means 
that their XRB populations are entirely comprised of low-mass X-ray binaries
(LMXBs).

Prior to the launch of \chandra\ only a small number of the very
brightest point-sources in early type galaxies could be 
resolved from the diffuse galactic emission
\citep[][]{fabbiano89}. \chandra's advent revolutionized this
picture, enabling LMXBs to be resolved in large numbers 
and studied directly \citep[\eg][for a recent review]{sarazin01,
blanton01b,humphrey04a,humphrey03a,fabbiano06a}.
Of particular interest are the X-ray luminosity functions (XLFs)
of the XRBs, which in principle may provide vital clues as 
to the way in which the sources form and evolve 
\citep[\eg][]{belczynski03a,ivanova06b}. In star-forming
galaxies, for example, the high-mass X-ray binary population 
dominates the XLF, making it much flatter than is typical
of LMXBs in the Milky Way or early-type galaxies
\citep{kilgard02,colbert04a,grimm03}.
Between early-type galaxies, the XLF appears remarkably 
similar \citep{kim04b,gilfanov04a}, although there is some
debate regarding its precise functional form. 
Various authors have parameterized it as a steeply falling 
broken powerlaw with a break around 2--4$\times 10^{38}$ \ergps\
and a high-luminosity negative logarithmic differential slope, 
$\beta \simeq$ 
2--3 \citep{sarazin00a,kraft01,colbert04a}, and there may be
evidence of a second break at $\sim 2\times 10^{37}$ \ergps 
\citep[\eg][]{gilfanov04a}. Alternatively, the XLF has also been modeled as a 
single powerlaw, truncated above some limit \citep{sivakoff03,jordan04a}. 
\citet{kim03a} argued
that the presence of a  break in the powerlaw XLF of NGC\thin 1316
can be artificially induced by source detection incompleteness
effects, after correcting for which the data were consistent
with a single powerlaw XLF. We obtained a similar result for the
lenticular galaxy NGC\thin 1332 \citep{humphrey04a}.
Applying an incompleteness correction to a sample
of $\sim$15 early-type galaxies, \citet{kim04b} also reproduced this
result, but they reported marginal evidence of a break at 
5$\times 10^{38}$ \ergps\ when the data for all the galaxies 
were combined.

Although the presence of the upper break is controversial, its luminosity
is suggestively close to the Eddington limit of a 1.4 ${\rm M_\odot}$ neutron star 
(${\rm L_{EDD}=}$2--4$\times 10^{38}$ \ergps, depending on the composition of 
the accreting matter and the neutron-star equation of state:
\citealt{paczynski83}), leading to suggestions that it may
delineate the division between neutron-star and black-hole
accretors \citep{sarazin01}. \citet{kim04b}, however, argued that
the break in their data was at too high a luminosity to make this explanation
tenable. Similarly, in the elliptical galaxy NGC\thin 720,
\citet{jeltema03} reported a break at ${\rm \sim 1\times 10^{39}}$ \ergps,
which they speculated may relate to a young stellar population
born in putative recent merger activity. 
\citet{bildsten04a} argued that if all bright LMXBs in early-type
galaxies are ultra-compact binaries with He or C/O-rich donor stars,
this might explain both the break and the low-luminosity slope of the 
XLF found by \citet{kim04b}.

Another issue for which studies of early-type galaxies are useful
is the role of globular
clusters (GCs) in forming LMXBs. It has long been recognized
that there is an over-density of LMXBs per unit optical
light  within Milky Way GCs as compared to the field, 
indicating that dynamical processes within them play an important
role in efficient LMXB formation \citep{fabian75a,clark75a}.
Since there are only $\sim$150 GCs in the Milky Way, however,
there are insufficient sources to examine this relationship in detail.
In contrast, many early-type galaxies host rich
GC populations \citep[\eg][]{gebhardt99a,kundu01b}, providing
ideal laboratories in which to investigate the LMXB-GC connection.
Using \asca\ data, \citet{white02a} found a strong correlation between the specific
frequency of globular clusters (\Sn) in a sample of 8 early-type
galaxies and the integrated luminosity of LMXBs, inferred 
from spectral-fitting, leading them to suggest that all LMXBs within
the galaxies were formed in GCs. With \chandra\ it has become
possible to isolate individual LMXBs associated with GCs and to 
study the populations of sources individually. The reported fraction of 
LMXBs associated with GCs varies from 
$\sim$20--70\%\  
\citep[\eg][]{angelini01,humphrey04a,kundu02a,xu05a,blanton01b}. 
Using a sample of 4  galaxies \citet{sarazin03} suggested this fraction
varies along the Hubble sequence, possibly reflecting the increase in
\Sn\ from spiral to elliptical galaxies, and perhaps indicating 
different populations of LMXBs which form in the field and in GCs.
\citet{irwin05a} and \citet{juett05a} reported further evidence that the 
fraction depends on \Sn.
\citeauthor{irwin05a} also argued 
that the  total luminosity of the LMXBs is not strictly proportional
to \Sn, which would imply a significant fraction of sources, even in 
early-type galaxies, forms in the field. 
The strength of the \citeauthor{juett05a} correlation, however, has been 
called into question by \citet{kim05a}. 

Another way to address this question is to compare the spatial distributions
of LMXBs and GCs, which has been attempted by \citet{kim05a} and 
\citet[][which was made publicly available after we submitted this present work]{kundu07a}, for small samples of galaxies (6 and 5, respectively). 
However, these studies have produced inconsistent results, which may 
reflect the small-number statistics of the samples. \citet{kim05a} 
found that the radial distribution of GCs hosting LMXBs is significantly
steeper than that of the GC population, but similar to the distribution
of LMXBs as a whole, and to the optical light. They argued this may indicate 
that dynamical processes affecting GCs close to the centres of each galaxy
may trigger LMXB production.  In contrast, \citet{kundu07a} found that the 
radial distribution of GCs hosting an LMXB is similar to the GC population
as a whole, and significantly flatter than the LMXB population, which
they argued hinted at significant formation in the field. 


Based on small samples of a handful of galaxies, it has been
shown that approximately 4\% of GCs are found to host active LMXBs 
\citep{angelini01,kundu02a,kundu03,sarazin03}.
Using a sample of 6 elliptical galaxies \citet{kim05a} 
found evidence that this fraction varies considerably from galaxy to galaxy.
There is no evidence that the properties of LMXBs in GCs
systematically differ from those in the field. In contrast,
GCs which are brighter or redder are systematically
more likely to contain an LMXB 
\citep{angelini01,kundu02a,kundu03,sarazin03,kim05a,smits06a,sivakoff06a}.
The luminosity dependence appears approximately consistent with the probability
that a GC contains an LMXB being exactly proportional to its
luminosity \citep[\eg][]{sarazin03,smits06a}, although \citet{sivakoff06a},
based on a sample of 11 Virgo galaxies, 
argued that the dependence upon mass is slightly stronger. 

The colour-dependence
is indicative of a metallicity effect \citep[\eg][]{kundu03},
and a number of authors have proposed possible explanations 
\citep[for a review, see][]{jordan04a}.
\citet{maccarone04a} proposed a model in which 
the effect arises due to the balance of mass transfer through 
irradiation-induced 
winds and Roche lobe overflow systematically varying from 
metal-poor to metal-rich systems. In this picture, LMXBs
associated with blue GCs should exhibit significant
intrinsic absorption. However, \citet{kim05a} reported no evidence of 
spectral differences between sources in red and in blue GCs.
Other possible explanations include the effect of metallicity on the 
stellar mass-radius relation \citep[][although this 
may not be sufficiently large an effect: \citealt{maccarone04a}]{bellazzini95a}, 
variation in the stellar initial mass function (IMF) with the GC metallicity
\citep[][although see \citealt{kroupa02a}]{grindlay87a} or the absence of a deep convective zone in metal-poor
stars impeding magnetic braking, and hence LMXB formation 
\citep[][although it is not clear if this model would produce the observed 
power-law dependence of LMXB incidence on the GC metallicity]{ivanova06c}.

To date the LMXB-GC connection has only been 
investigated with small galaxy samples, making it hard to draw strong 
conclusions about the relation between them in general. In particular
the fraction of LMXBs which may be formed in the field is very uncertain.
In addition, the presence of a break in the XLF remains controversial. 
The investigation
of the LMXBs in a larger sample of galaxies is therefore vital to
provide a clearer insight into the processes which give rise to them.
In this paper, we investigate the LMXB population of  24 early-type galaxies 
observed with \chandra\ and \hst, focusing on  both the XLF and the LMXB-GC 
connection.
This work is also part of a series investigating the X-ray properties
of a sample of galaxies. In our previous papers we have addressed the
metal content of the ISM and their mass profiles
\citep{humphrey05a,humphrey06a}. We consider here only a subset of 
the galaxies which yield interesting
constraints on the LMXB populations. In addition, we include the nearby
galaxy NGC\thin 1404, the diffuse emission of which is difficult to 
disentangle from that of the Fornax ``cluster'' but which has 
a fairly rich LMXB and GC population. The galaxies and a 
summary of their properties, and the \chandra\ observations used,
are listed in Table~\ref{table_obs}.
In a companion paper,
\citet[][hereafter \gcpaper]{humphrey06c}, we present the \hst\
data-reduction and analysis upon which we build here.
Of the 24 galaxies in the sample, 7 were included in the \citet{kim04b}
study of the XLF, 3 in the sample of \citet{kim05a}, who considered
the LMXB-GC connection, 4  in the \citet{smits06a} sample, 
and 3 in the \citet{sarazin03}
sample. All errors quoted in this paper are 90\% confidence limits unless 
otherwise stated. 

\section{Observations and data analysis}
\pagestyle{empty}
\renewcommand{\tabcolsep}{2mm}
\begin{deluxetable*}{llllllllllll}
\tablewidth{0pt}
\tablecaption{Target list and observation log \label{table_obs}}
\tablehead{
\colhead{Name}&\colhead{Type}&\colhead{Dist}&\colhead{\dtwentyfive}&\colhead{$L_K$}& \colhead{Age}&\colhead{[Fe/H]}&\colhead{[$\alpha$/Fe]}&\colhead{Obsid}&\colhead{Inst}&\colhead{Date}&\colhead{Exp} \\
\colhead{ }&\colhead{ }&\colhead{Mpc }&\colhead{\arcmin}&\colhead{${\rm 10^{11} L_\odot}$ }&\colhead{Gyr }&\colhead{ }&\colhead{ }&\colhead{ }&\colhead{ }&\colhead{dd/mm/yy }&\colhead{ks }}
\startdata
IC4296 &E-Radio-gal &$50.8$ &$3.8$ &$5.2$ &$12.$ &$-0.10\pm 0.07$ &$0.31\pm 0.05$ &3394 &S &10/09/01 &$25.$ \\
NGC720 &E5 &$25.7$ &$4.6$ &$1.7$ &$2.9^{+0.8}_{-0.2}$ &$0.17\pm 0.13$ &$0.37\pm 0.03$ &492 &S &12/10/00 &$29.$ \\
NGC1332 &S(s)0 &$21.3$ &$4.1$ &$1.4$ &$4.1^{+5.4}_{-0.9}$ &$-0.15\pm 0.14$ &$0.31\pm 0.10$ &4372 &S &19/09/02 &$45.$ \\
NGC1387 &SAB(s)0 &$18.9$ &$3.3$ &$0.78$ &\ldots &\ldots &\ldots &4168 &I &20/05/03 &$45.$ \\
NGC1399 &cD;E1pec &$18.5$ &$6.9$ &$2.1$ &$12.\pm 2.$ &$-0.11\pm 0.08$ &$0.37\pm 0.03$ &319 &S &18/01/00 &$56.$ \\
NGC1404 &E1 &$19.5$ &$4.1$ &$1.5$ &$12.^{+2.}_{-3.}$ &$-0.28\pm 0.09$ &$0.25\pm 0.03$ &2942 &S &13/02/03 &$29.$ \\
NGC1407 &E0 &$26.8$ &$5.3$ &$3.1$ &$12.\pm 1.$ &$-0.18\pm 0.05$ &$0.33\pm 0.01$ &791 &S &16/08/00 &$40.$ \\
NGC1549 &E0-1 &$18.3$ &$4.7$ &$1.3$ &$5.1\pm 0.4$ &$-0.10\pm 0.04$ &$0.240\pm 0.006$ &2077 &S &08/11/00 &$22.$ \\
NGC1553 &SA(rl)0;LINER &$17.2$ &$5.3$ &$1.9$ &$5.7^{+0.7}_{-0.5}$ &$0.23\pm 0.02$ &$0.170\pm 0.006$ &783 &S &02/01/00 &$14.$ \\
NGC3115 &S0 &$9.00$ &$7.3$ &$0.74$ &$15.\pm 3.$ &$0.030\pm 0.10$ &$0.11\pm 0.07$ &2040 &S &14/06/01 &$36.$ \\
NGC3585 &E7/S0 &$18.6$ &$6.1$ &$1.5$ &$10.^{+3.}_{-4.}$ &$-0.13\pm 0.11$ &$0.20\pm 0.07$ &2078 &S &03/06/01 &$35.$ \\
NGC3607 &SA(s)0 &$21.2$ &$4.5$ &$1.5$ &$15.\pm 2.$ &$-0.13\pm 0.11$ &$0.19\pm 0.07$ &2073 &I &12/06/01 &$38.$ \\
NGC3923 &E4-5 &$21.3$ &$6.4$ &$2.3$ &$3.3^{+0.4}_{-0.2}$ &$0.13\pm 0.06$ &$0.34\pm 0.02$ &1563 &S &14/06/01 &$8.8$ \\
NGC4125 &E6pec;Liner &$22.2$ &5.8 &$1.8$ &$13.\pm 5.$ &$-0.39\pm 0.17$ &$0.33\pm 0.10$ &2071 &S &09/09/01 &$63.$ \\
NGC4261 &E2-3;Liner;Sy3 &$29.3$ &4.1 &$2.2$ &$15.0\pm 0.6$ &$-0.21\pm 0.06$ &$0.25\pm 0.01$ &834 &S &06/05/00 &$34.$ \\
NGC4365 &E3 &$19.0$ &$5.8$ &$1.6$ &$3.9^{+5.9}_{-0.7}$ &$-0.020\pm 0.12$ &$0.19\pm 0.09$ &2015 &S &02/06/01 &$40.$ \\
NGC4472 &E2/S0(2);Sy2 &$15.1$ &$9.7$ &$3.2$ &$9.0\pm 1.2$ &$-0.0100\pm 0.12$ &$0.16\pm 0.02$ &321 &S &12/06/00 &$32.$ \\
NGC4494 &E1-2;LINER &$15.8$ &$4.5$ &$0.81$ &$15.\pm 5.$ &$-0.24\pm 0.12$ &$0.16\pm 0.08$ &2079 &S &05/08/01 &$15.$ \\
NGC4552 &E;LINER-HII &$14.3$ &$5.0$ &$0.85$ &$12.\pm 1.$ &$-0.050\pm 0.061$ &$0.24\pm 0.01$ &2072 &S &22/04/01 &$54.$ \\
NGC4621 &E5 &$17.0$ &$5.0$ &$1.2$ &$7.1^{+4.8}_{-2.3}$ &$-0.030\pm 0.11$ &$0.28\pm 0.09$ &2068 &S &01/08/01 &$25.$ \\
NGC4649 &E2 &$15.6$ &7.4 &$2.4$ &$13.\pm 1.$ &$0.050\pm 0.085$ &$0.25\pm 0.01$ &785 &S &20/04/00 &$21.$ \\
NGC5018 &E3 &$42.6$ &$3.6$ &$3.0$ &$2.0^{+1.4}_{-0.2}$ &$0.15\pm 0.17$ &$0.0100\pm 0.085$ &2070 &S &14/04/01 &$28.$ \\
NGC5845 &E &$24.0$ &$0.90$ &$0.27$ &$12.$ &$-0.12\pm 0.16$ &$0.26\pm 0.10$ &4009 &S &03/01/03 &$30.$ \\
NGC5846 &E0-1;LINER-HII &$21.1$ &$3.8$ &$1.5$ &$15.\pm 1.$ &$-0.18\pm 0.10$ &$0.22\pm 0.01$ &788 &S &24/05/00 &$23.$ \\
\enddata
\tablecomments{Target list and observation details. All distances (Dist) are  taken 
from  \citet{tonry01}, corrected for the the new Cepheid zero-point \citep{jensen03},
except for IC\thin 4296, taken from \citet{jensen03}, and NGC\thin 5018
from ${\rm D_n}$-$\sigma$ relation
\citep{faber89}. B-band twenty-fifth magnitude isophote (\dtwentyfive) 
diameters are taken from RC3.
K-band luminosities (\lk) were taken from \twomass\ \citep{jarrett00a}, 
adopting $M_{K\odot}=3.41$. We report the age of the stellar population, 
its metallicity, [Fe/H] and its $\alpha$-to-Fe ratio, [$\alpha$/Fe] 
(see \S\ref{sect_slope}), and their 1-$\sigma$ errors.
The \chandra\ observation identifier is given
(ObsID), as is the ACIS instrument (inst), the date of the observation start and
the total exposure (exp), having cleaned the data to remove flaring (see text).}
\end{deluxetable*}

For data reduction, we used the \ciao\ 3.3.0.1 and \heasoft\ 6.0 software
suites, in conjunction with \chandra\ calibration database (\caldb)
version 3.2.1. Spectral-fitting was conducted with \xspec\ 11.3.2p.
In order to ensure the most up-to-date calibration, all data were 
reprocessed from the ``level 1'' events files, following the standard
\chandra\ data-reduction threads\footnote{{http://cxc.harvard.edu/ciao/threads/index.html}}.
Bad-pixel maps were created using the \ciao\ tool {\tt acis\_run\_hotpix}.
We applied the standard correction to take account of the time-dependent 
gain-drift and, for those galaxies observed with ACIS-I, we applied the 
standard CTI correction
as implemented in the standard \ciao\ tools. 

To identify periods of enhanced
background (``flaring''), which seriously degrades the signal-to-noise (S/N)
we accumulated background lightcurves for each exposure from
low surface-brightness regions of the active chips. We
excluded obvious diffuse emission and data in the vicinity of any detected
point-sources (see below). Periods of flaring were identified by eye and
excised. The final exposure times are listed in Table~\ref{table_obs}.

Point source detection was performed using the \ciao\ tool
{\tt wavdetect} \citep{freeman02}. Point sources were identified 
in full-resolution images of the \acis\ focal-plane containing,
where appropriate, the S3 chip and 
any other chips onto which the 
B-band twenty-fifth magnitude (\dtwentyfive) ellipse
listed in the Third Catalogue of Bright Galaxies 
\citep[hereafter RC3]{devaucouleurs91} extends.
For ACIS-I observations, all of the ACIS-I chips were considered.
To maximise the likelihood of identifying
sources with peculiarly hard or soft spectra, images were created in three
energy bands, 0.1--10.0~keV, 0.1--3.0~keV and 3.0--10.0~keV. Sources were
detected separately in each image. In order to minimize spurious detections at
node or chip boundaries we supplied the detection algorithm with
exposure-maps  generated at
energies 1.7~keV, 1.0~keV and 7~keV respectively (although the precise
energies chosen should make little difference to the results). The
detection algorithm searched for structure over pixel-scales of 1, 2, 4, 8 and
16 pixels, and the detection threshold was set to $10^{-6}$, implying
$\sim$1 spurious detection due to background fluctuations per CCD. 
The source-lists obtained in each energy-band were combined and
duplicated sources removed, and the final list was checked
by visual inspection of the images. For each source, we obtained
radial surface brightness profiles, as described in \citet{humphrey04a},
which we examined to ensure that only point-like sources were 
considered (where there were sufficient photons). 
This led to the elimination of central sources in many of 
the galaxies, which were spuriously detected by the algorithm due to
a centrally-peaked surface brightness profile.
In a few other cases (notably in the centre of NGC\thin 5846), other
``sources'' eliminated in this way appeared to be associated with compact
knots of X-ray emission, or the sharp edges of X-ray structures.
The spiral galaxy NGC\thin 4647 is projected onto the part of the 
image of NGC\thin 4649, possibily contaminating our source list for this 
object. In order to minimize such contamination, we excluded all point-sources
which lay within the \dtwentyfive\ ellipse of NGC\thin 4647 from further
analysis.

Spectra were extracted for each source from the 3-$\sigma$ source regions
returned by the detection algorithm. We note that there was very little problem
of these regions overlapping; in the few cases where they were found to overlap,
the regions were slightly shrunk to minimize likelihood of  contamination, but this
should not strongly affect the measured fluxes.
For each source spectra and spectral responses
were generated using the standard \ciao\ tools and
local background regions were chosen
from which background products were similarly extracted. 
The background regions were chosen to cover an area at least 8 times the 
source region, and containing at least 50 photons, and excluding data from the 
6-$\sigma$ detection region of {\em all} detected sources. Furthermore,
we restricted the background regions to lie  entirely on the same
chip and CCD node as the source centroid. A few sources were eliminated at this
stage as they contained no (background-subtracted) photons; 
these are most likely spurious detections due to background fluctuations. 
Complete lists of the detected sources
which lie within \dtwentyfive\ are given in the Appendix.

\section{X-ray source properties}
\subsection{Hardness ratios} \label{sect_hardness}
\begin{figure*}
\centering
\includegraphics[scale=0.7]{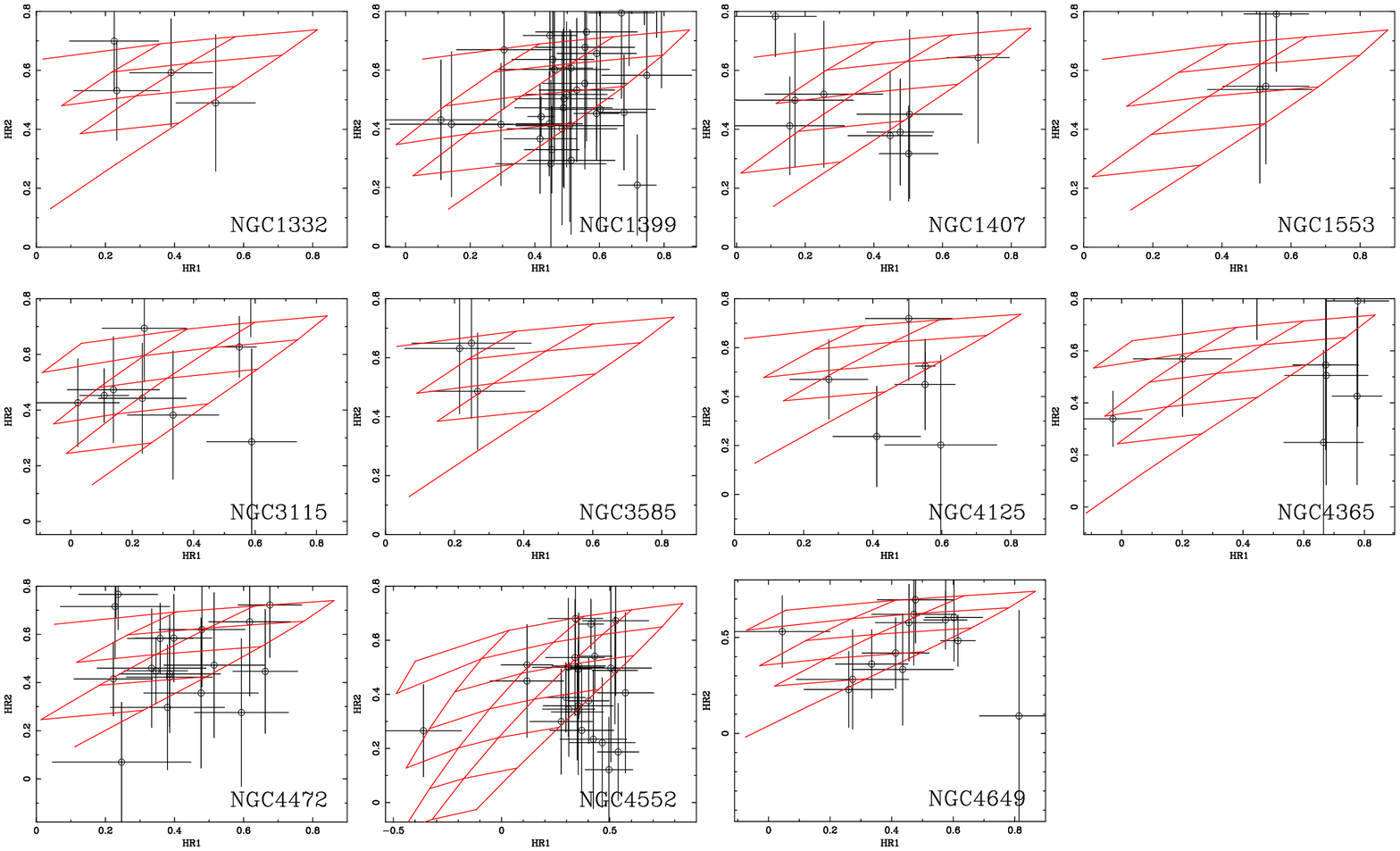}
\caption{Hardness ratio plots of point sources with more than 50 photons
for a subsample of the galaxies. The definitions of the hardness ratios
HR1 and HR2 are given in the text. Also shown is a grid of loci in the 
colour-colour plane for a simple photo-absorbed powerlaw as \nh\ and 
$\Gamma$ are allowed to vary. \nh\ increases to the left and grid lines are 
shown for \nh=0.0, 0.25, 0.5, 1.0, 2.0, 4.0 and 
8.0$\times 10^{22}{\rm cm^{-2}}$. 
$\Gamma$ increases upwards, with grid lines shown for 
$\Gamma=$0.0--3.0, in steps of 0.5} \label{fig_hardness}
\end{figure*}
Individual point-sources typically contained too few background-subtracted
photons to enable full spectral-fitting. Therefore to obtain a crude
insight into their spectral properties, we computed spectral hardness
ratios \citep[see \eg][]{kilgard02,humphrey03a}. We defined 
HR1 and HR2 for each source as, respectively, HR1=(S-M)/(S+M)
and HR2=(M-H)/(M+H), where S is the background subtracted count-rate
in the band 0.3--1.5~keV, M is the count-rate in the band 
1.5--3.0~keV and H corresponds to 3.0--5.0~keV.
Fig~\ref{fig_hardness} shows, for a subsample of the galaxies,
plots of the hardness
ratios for those sources with more than 50 photons. 
For comparison purposes, we also show the loci in the 
HR2--HR1 plane of a simple absorbed powerlaw as \nh\ and $\Gamma$ 
are allowed to vary. The exact positions of these grids do not vary
substantially between observations.

Most of the sources are consistent with having low
column densities (for typical Galactic line-of-sight \nh\ values
the grid lines are practically indistinguishable from the \nh=0 case),
and powerlaw slopes, $\Gamma \sim$1--3. Such parameters are
as expected for an LMXB population in an early-type galaxy 
\citep[see \eg][]{humphrey03a}.
A small number of sources exhibit much harder colours, implying 
intrinsic absorption. Some or all of these objects may be 
heavily-absorbed background AGN. We do not exclude such objects
from subsequent analysis, since we explicitly take account of
the expected numbers of background sources.

\subsection{Fluxes} \label{sect_fluxes}
Since direct spectral fitting was impractical, it was necessary to
adopt a canonical model for the spectrum to obtain a flux estimate 
for each source. We adopted a simple powerlaw model, with 
$\Gamma=1.55$, which has been shown to fit adequately the 
composite spectra of detected LMXB 
\citep[][see also \S~\ref{sect_composite_spectra}]{irwin03a},
and modified by photoelectric absorption due to the Galactic ISM along
the  line-of-sight \citep{dickey90}. Such a model is broadly
consistent with the source hardness ratios (\S~\ref{sect_hardness})
and models of a similar shape have also been fitted to the spectra of 
LMXBs within the Milky Way \citep[\eg][]{church01}.
This model was folded through the response matrices generated 
at each source position, and used to infer a counts-to-flux
conversion ratio in the 0.3--7.0~keV band. The 0.3--7.0~keV 
background-subtracted fluxes of each source were extracted from the regions 
defined from which to extract spectra.
Although a small number of sources appear heavily absorbed 
(\S~\ref{sect_hardness}), and so this count-to-flux conversion ratio
is not formally correct, the X-ray luminosity function we adopt
to account for interlopers (\S~\ref{sect_xlf}) 
similarly does not incorporate the intrinsic
absorption in estimating the flux, and so the results are self-consistent.
Given the narrow band in which we compute fluxes, 
by definition, they underestimate the bolometric 
flux of the source. However, the extrapolation of the spectrum outside
the adopted band is highly uncertain and so we compute the XLF of 
the sources for this narrow band.
The luminosities are shown for each source in the Appendix.

\subsection{Composite spectra} \label{sect_composite_spectra}
\begin{figure}
\centering
\includegraphics[scale=0.35]{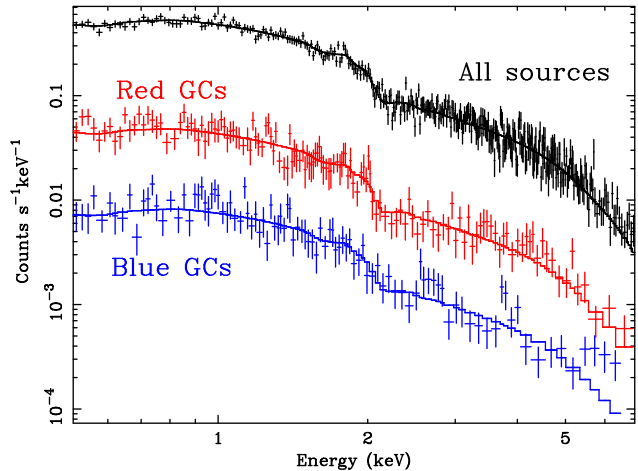}
\caption{Composite spectra of the LMXBs (upper spectrum; black),
folded through the instrumental response. The data are shown fitted
with an absorbed bremsstrahlung model, with kT fixed at 7.3~keV.
Also shown are the composite spectrum of LMXBs associated with 
red GCs and, scaled for clarity by a factor 0.3, that of LMXBs
associated with blue GCs (\S \ref{sect_gc_colours}). 
These spectra are shown fitted
with the same bremsstrahlung model.} \label{fig_spectra}
\end{figure}
With a sample of 15 galaxies, \citet{irwin03a} showed that the composite
spectrum of all the detected LMXBs is featureless and 
can be well-fitted by simple empirical models
such as a powerlaw with $\Gamma=$1.56 or thermal bremsstrahlung with
kT=7.3~keV. In order to test this with our sample, for each galaxy, we 
accumulated composite source and background spectra for all of the 
observed sources which lie within \dtwentyfive,
excluding the central $\sim$20\arcsec\ (where background subtraction
may be more problematical). These spectra were added in count-rate
space, having first scaled the background spectra by the ratio of the 
source to background BACKSCAL keywords.
The response matrices were averaged together with the \heasoft\
tools {\tt addrmf} and {\tt addarf} and adopting the 
background-subtracted count-rate of each corresponding spectrum
as a weighting factor.

We fitted the composite spectrum in the range 0.5--7.0~keV, having
first rebinned it to ensure S/N of 3 in each bin, and at least 20
photons (to allow the use of the $\chi^2$ fitting statistic). We 
fitted the spectrum using Xspec.
The spectrum is remarkably featureless (Fig~\ref{fig_spectra})
and well-fitted by a single, absorbed powerlaw ($\chi^2$/dof=354/339),
with $\Gamma=1.68\pm0.04$ and \nh=${\rm (7.5\pm1.1)\times 10^{20} cm^{-2}}$.
If we fitted the data, instead, with a bremsstrahlung model,
we obtained a similarly good fit ($\chi^2$/dof=357/339), with
kT=$7.8^{+0.8}_{-0.7}$ and \nh=${\rm (2.1\pm0.08)\times 10^{20} cm^{-2}}$.
These results are consistent with \citet{irwin03a}, although
$\Gamma$ is slightly larger for the powerlaw case.
It is interesting to note that \nh\ obtained for the bremsstrahlung
case is in better agreement with the average Galactic line-of-sight 
column density for the galaxies in our sample (${\rm \sim 3\times 10^{20} cm^{-2}}$) 
\citep[see][]{humphrey05a}, implying it is more representative.
Fitting this model to the composite source spectra accumulated for
each galaxy individually, with \nh\ fixed to the nominal value 
for the appropriate Galactic line-of-sight, we obtained acceptable
fits for all of the galaxies.

\subsection{Spatial distribution} \label{sect_spatdist}
\begin{figure*}
\centering
\includegraphics[scale=0.8]{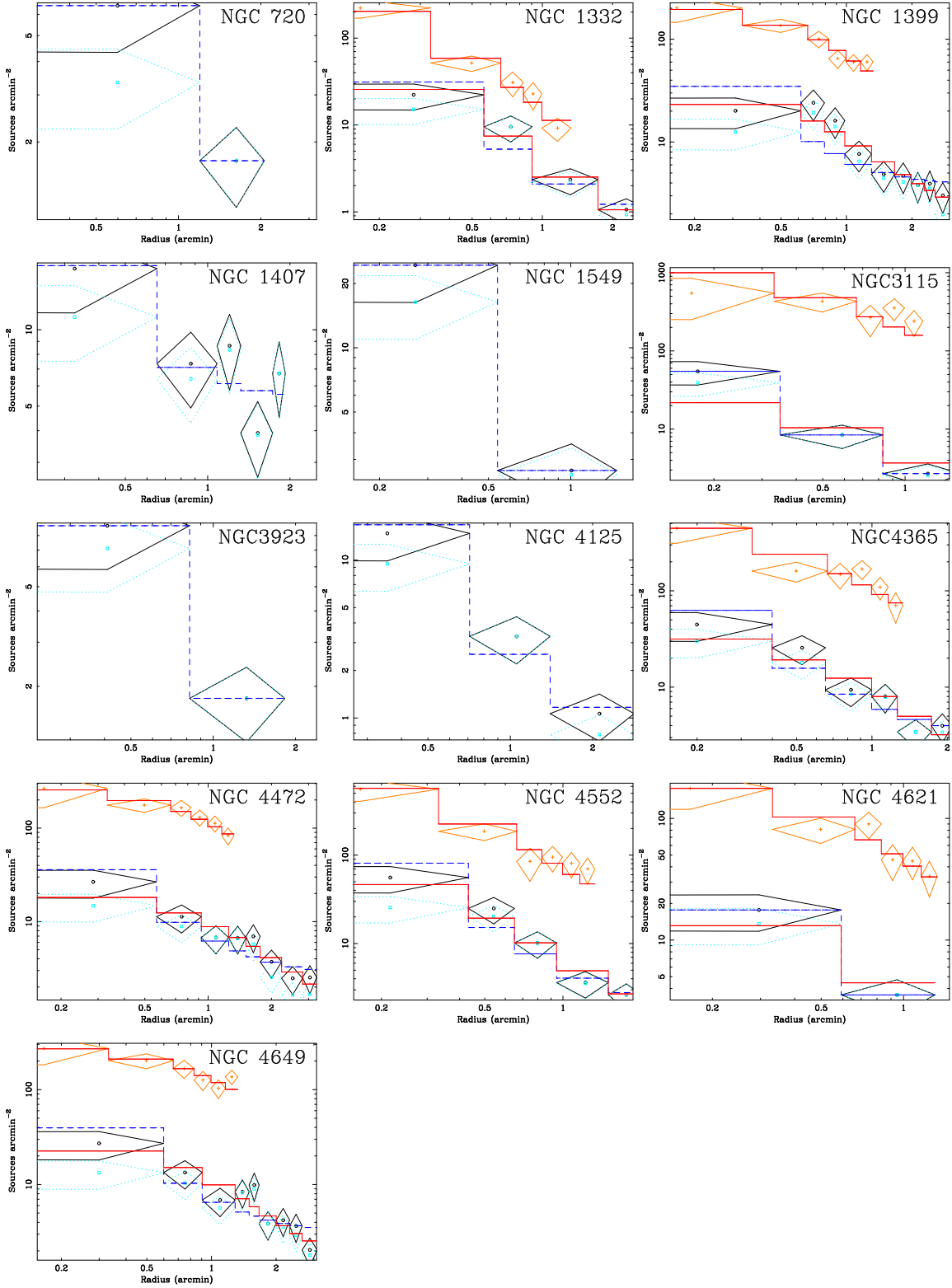}
\caption{Spatial distribution of the LMXBs in those galaxies with
sufficient sources. Shown are the data-points having corrected for spatial
variation in source detection incompleteness (circles; black) and, where
appropriate, the best-fitting model to a suitably weighted combination
of the distributions of red and blue GCs
(solid line; red). The distribution of the GCs, corrected for 
incompleteness, is also shown, with arbitrary scaling for clarity 
(crosses; orange) and the same best-fitting model.
The dashed line (blue)
shows the distribution of the optical light.
For comparison, we also show the data-points
having not corrected for the spatial variation in source incompleteness 
(triangles; light blue).} \label{fig_spatdist}
\end{figure*}
\begin{deluxetable}{llll}
\tablecaption{LMXB spatial distribution goodness-of-fit\label{table_spatprob}}
\tablehead{
\colhead{Galaxy} & \colhead{prob(light)} & \colhead{prob(GC,tot)} &
\colhead{prob(GC,weight)}
}
\startdata
NGC720 &$0.35$ &\ldots &\ldots \\
NGC1332 &$0.10$ &$0.15$ &$0.090$ \\
NGC1399 &$1.0\times 10^{-3}$ &$0.48$ &$0.85$ \\
NGC1407 &$0.20$ &\ldots &\ldots \\
NGC1549 &$0.12$ &\ldots &\ldots \\
NGC3115 &$0.78$ &$0.010$ &$0.060$ \\
NGC3923 &$0.20$ &\ldots &\ldots \\
NGC4125 &$0.14$ &\ldots &\ldots \\
NGC4365 &$0.070$ &$0.070$ &$0.22$ \\
NGC4472 &$0.23$ &$0.29$ &$0.97$ \\
NGC4552 &$0.080$ &$0.010$ &$0.44$ \\
NGC4621 &$0.24$ &$0.13$ &$0.21$ \\
NGC4649 &$3.0\times 10^{-3}$ &$0.32$ &$0.44$ \\
\enddata
\tablecomments{The probability that the radial distribution of LMXBs 
is the same of the optical light 
(prob(light)) in each galaxy and the probability that 
it is the same as that of the GC population (prob(GC,tot)), or
a weighted combination
of the red and blue GC distributions (prob(GC,weight); {\em see text}).
}
\end{deluxetable}
We next investigated the spatial distribution of the point-sources in
a subset of the galaxies which contained sufficient sources. We created
a set of radial bins centred on the X-ray centroid and  each containing 
at least 15 sources. We compared the radial profiles of the sources
with the optical light, which we modeled as a de Vaucouleurs profile,
the effective radius of which was fixed to the appropriate K-band value 
determined from 2MASS \citep{jarrett00a}. We also included an additional,
constant, component to account for interloper sources, which should
be approximately uniformly sprinkled over the \chandra\ field. Fitting 
was performed with a Cash-C statistic mimimization algorithm.

Since the effects of source detection incompleteness
depend on the density of sources, the width of the point-spread function
and the ``background'' count-rate \citep[\eg][]{kim03a}, it varies
spatially across the field of view. It was therefore necessary to
take this into account when fitting the radial profile of sources.
We discuss in general how we estimated the effects of source 
detection incompleteness on the XLF
in \S~\ref{sect_incompleteness}.  We computed a similar correction
appropriate for each radial bin, with which we weighted an assumed
X-ray luminosity function (assumed to the best-fitting broken powerlaw
XLF we found in \S~\ref{sect_xlf}) that
was then integrated from $10^{37}$--$2\times 10^{39}$\ergps\ 
so as to estimate the 
fraction of sources detected as a function of radius. 

In Fig~\ref{fig_spatdist} we show the spatial distribution of the
point-sources in the subsample, both with 
a correction for the spatial variation in completeness applied, 
and with no correction. Outside the
innermost 0.5\arcmin\ (in which  incompleteness effects are most severe)
the radial distribution of sources agrees very well with the optical
light. Inside this region we find an excellent agreement between
the source distribution and the optical light in approximately
half of the tested galaxies. In the remaining galaxies, the optical
light appears to over-predict the number of sources considerably.
To determine the goodness-of-fit, we performed 100 Monte-Carlo simulations
(increasing the number if the estimated goodness-of-fit was below 1\%),
where artificial data-sets were simulated from the best-fitting models
and were fitted with the same model. 
The fraction of simulations with a larger Cash-C statistic (\ie\ poorer
fit) than for the real data was adopted as the null hypothesis
probability (prob(light)), as shown in Table~\ref{table_spatprob}.
If the LMXBs are, in general, distributed like the optical light, 
the distribution of prob(light) should, for these 13 galaxies, be 
uniform over the interval 0--1. Using a Kolmogorov-Smirnov (K-S) test
\citep[as implemented in][]{nr}, the 
probability the distributions are the same is only $2\times 10^{-5}$. 
Considering only those galaxies for which we also have sufficient-quality
GC data to measure the GC radial distribution as well (below), 
this probability is 0.05\%.

If a substantial fraction of LMXBs are produced in GCs, we might expect
the radial distributions of the two populations to be similar. 
Since the field of view 
of the WFPC2 is significantly more restricted than \chandra, it was
necessary to fit the LMXB and GC data with the same model simultaneously. 
To parameterize the GC distribution, we adopted a ``beta'' model 
($\propto (1+(R/R_c)^2)^{0.5-3\epsilon}$,
where $R_c$ and $\epsilon$ are parameters of the fit), and we 
additionally allowed a constant background component for the LMXBs
(as discussed above). The data were of sufficient quality to enable
this for 8 of the galaxies, and the null hypothesis probabilities
are reported for this comparison in Table~\ref{table_spatprob}.
We find moderately good 
agreement in most of the galaxies between the two distributions, although
in two galaxies the null hypothesis probability was $<$5\%. 
Blue GCs are known to have a considerably
flatter distribution than their red counterparts (\eg\ \gcpaper), while LMXBs
are preferentially found in red GCs, and so a comparison with the radial
distribution of the total GC population may not be entirely appropriate.
We therefore also compared the radial 
distribution of the LMXBs to a weighted combination of the radial profiles of 
the red and blue GCs, taking into account the fraction of LMXBs 
expected to be produced in each population. 
For this model we found reasonably good agreement between the 
distributions in all of the galaxies.  The null hypothesis probabilities for 
these fits are shown in Table~\ref{table_spatprob}. 
Based on a K-S test, as described above, we found that the distributions of 
these probabilities agrees with expectation (a uniform distribution) with
a null hypothesis probability of 17\%, significantly higher than the equivalent
comparison with the optical light (0.05\%).

\section{X-ray luminosity functions}
\renewcommand{\tabcolsep}{0.3mm}

\begin{deluxetable*}{lllllllllllll}
\centering
\tabletypesize{\tiny}
\tablecaption{XLF fitting results \label{tab_xlf}}
\tablehead{
\colhead{Galaxy }&\colhead{$N_{src}$}&\colhead{$f_{cover}$}&\colhead{$L_{X,min}$}&\colhead{${L_{X,20}}$}&\colhead{Prob }&\colhead{beta }&\colhead{\lx}&\colhead{\probhnought}&\colhead{$\beta_1$ }&\colhead{$L_{break}$}&\colhead{$\beta_2$}&\colhead{\lx} \\
\colhead{ }&\colhead{ }&\colhead{ }&\colhead{($10^{38}$}&\colhead{($10^{38}$}&\colhead{ }&\colhead{ }&\colhead{($10^{38}$}&\colhead{}&\colhead{ }&\colhead{($10^{38}$}&\colhead{ }&\colhead{($10^{38}$}\\
\colhead{ }&\colhead{ }&\colhead{ }&\colhead{\ergps)}&\colhead{\ergps)}&\colhead{ }&\colhead{ }&\colhead{\ergps)}&\colhead{}&\colhead{ }&\colhead{\ergps)}&\colhead{ }&\colhead{\ergps)}
}
\startdata
IC4296 &$11$ &$0.41$ &$8.1$ &$20.$ &$0.080$ &$2.9^{+1.5}_{-0.9}$ &$8200^{+1400000}_{-7800}$ &$0.16$ &$1.4$ &$2.2$ &\ldots &$940^{+730}_{-520}$ \\
NGC720 &$22$ &$0.37$ &$1.0$ &$4.8$ &$0.050$ &$2.1\pm 0.4$ &$400^{+300}_{-170}$ &$0.13$ &$1.4$ &$2.2$ &$<$$2.2$ &$280\pm 130$ \\
NGC1332 &$29$ &$0.25$ &$0.47$ &$1.5$ &$0.69$ &$2.4^{+0.5}_{-0.4}$ &$320\pm 150$ &$0.49$ &$1.4$ &$2.2$ &\ldots &$240^{+90}_{-70}$ \\
NGC1387 &$10$ &$0.38$ &$0.96$ &$1.9$ &$0.50$ &$3.5\pm 1.5$ &$370^{+1700}_{-290}$ &$0.060$ &$1.4$ &$2.2$ &\ldots &$61.^{+48.}_{-34.}$ \\
NGC1399 &$144$ &$0.57$ &$0.22$ &$0.91$ &$0.11$ &$2.2\pm 0.2$ &$690\pm 110$ &$<0.01$ &$1.4$ &$2.2$ &$3.2^{+0.9}_{-0.6}$ &$540\pm 90$ \\
NGC1404 &$19$ &$0.40$ &$1.1$ &$2.1$ &$0.040$ &$2.3^{+0.5}_{-0.4}$ &$230\pm 190$ &$0.61$ &$1.4$ &$2.2$ &$>$$1.5$ &$140^{+70}_{-50}$ \\
NGC1407 &$88$ &$0.48$ &$0.71$ &$2.6$ &$0.13$ &$2.0\pm 0.2$ &$630\pm 140$ &$0.39$ &$1.4$ &$2.2$ &$2.3^{+0.6}_{-0.5}$ &$470^{+100}_{-90}$ \\
NGC1549 &$37$ &$0.49$ &$0.53$ &$2.4$ &$0.51$ &$2.1\pm 0.3$ &$160^{+60}_{-50}$ &$0.51$ &$1.4$ &$2.2$ &$>$$1.4$ &$140^{+50}_{-40}$ \\
NGC1553 &$23$ &$0.39$ &$0.82$ &$1.9$ &$0.010$ &$1.8\pm 0.4$ &$100^{+60}_{-40}$ &$0.26$ &$1.4$ &$2.2$ &\ldots &$82.^{+44.}_{-34.}$ \\
NGC3115 &$44$ &$0.37$ &$0.091$ &$0.36$ &$0.13$ &$1.7\pm 0.2$ &$79.^{+53.}_{-36.}$ &$0.60$ &$1.4$ &$2.2$ &\ldots &$66.\pm 23.$ \\
NGC3585 &$31$ &$0.53$ &$0.38$ &$1.7$ &$0.20$ &$2.0^{+0.4}_{-0.3}$ &$110^{+50}_{-40}$ &$0.47$ &$1.4$ &$2.2$ &\ldots &$97.\pm 38.$ \\
NGC3607 &$16$ &$0.41$ &$0.85$ &$2.9$ &$0.36$ &$3.5^{+1.4}_{-0.9}$ &$670^{+3900}_{-500}$ &$0.030$ &$1.4$ &$2.2$ &\ldots &$94.^{+56.}_{-43.}$ \\
NGC3923 &$30$ &$0.51$ &$0.83$ &$4.4$ &$0.56$ &$2.2^{+0.5}_{-0.4}$ &$240^{+200}_{-100}$ &$0.49$ &$1.4$ &$2.2$ &\ldots &$150^{+70}_{-60}$ \\
NGC4125 &$31$ &$0.45$ &$0.47$ &$1.3$ &$0.040$ &$1.9^{+0.5}_{-0.4}$ &$88.^{+56.}_{-42.}$ &$0.19$ &$1.4$ &$2.2$ &$<$$4.6$ &$74.^{+41.}_{-33.}$ \\
NGC4261 &$38$ &$0.49$ &$0.69$ &$3.7$ &$0.070$ &$2.7^{+0.5}_{-0.4}$ &$610^{+790}_{-280}$ &$0.16$ &$1.4$ &$2.2$ &$>$$1.7$ &$240^{+90}_{-70}$ \\
NGC4365 &$104$ &$0.53$ &$0.30$ &$1.4$ &$0.37$ &$2.0\pm 0.2$ &$320^{+70}_{-60}$ &$0.59$ &$1.4$ &$2.2$ &$3.0\pm 1.6$ &$290\pm 60$ \\
NGC4472 &$145$ &$0.61$ &$0.22$ &$1.2$ &$0.050$ &$2.2\pm 0.2$ &$470\pm 80$ &$<0.01$ &$1.4$ &$2.2$ &$3.0^{+1.2}_{-0.9}$ &$410\pm 70$ \\
NGC4494 &$13$ &$0.46$ &$0.82$ &$2.6$ &$0.60$ &$2.2^{+1.1}_{-0.6}$ &$58.^{+60.}_{-34.}$ &$0.91$ &$1.4$ &$2.2$ &\ldots &$47.^{+37.}_{-27.}$ \\
NGC4552 &$79$ &$0.49$ &$0.13$ &$0.61$ &$0.13$ &$1.8\pm 0.2$ &$180^{+60}_{-50}$ &$0.81$ &$1.4$ &$2.2$ &$>$$0.74$ &$160^{+40}_{-30}$ \\
NGC4621 &$34$ &$0.49$ &$0.45$ &$1.6$ &$0.24$ &$2.0^{+0.4}_{-0.3}$ &$87.^{+47.}_{-35.}$ &$0.70$ &$1.4$ &$2.2$ &$<$$4.2$ &$80.\pm 37.$ \\
NGC4649 &$121$ &$0.59$ &$0.25$ &$1.4$ &$0.13$ &$2.1\pm 0.2$ &$420^{+80}_{-70}$ &$0.17$ &$1.4$ &$2.2$ &$2.7^{+1.0}_{-0.8}$ &$370^{+70}_{-60}$ \\
NGC5018 &$13$ &$0.46$ &$2.4$ &$9.5$ &$0.57$ &$2.3^{+1.5}_{-0.7}$ &$330^{+3500}_{-230}$ &$0.73$ &$1.4$ &$2.2$ &$>$$1.2$ &$170\pm 140$ \\
NGC5846 &$16$ &$0.45$ &$1.7$ &$3.5$ &$0.22$ &$2.4^{+0.9}_{-0.6}$ &$240^{+920}_{-150}$ &$0.47$ &$1.4$ &$2.2$ &$>$$0.54$ &$110^{+70}_{-50}$ \\
\hline
Composite & \ldots & \ldots & 0.1 & \ldots &  $<$0.001 &  1.84$\pm0.05$  & 1950$\pm$140 & 0.03 & 1.40$^{+0.11}_{-0.13}$ &2.2$^{+0.65}_{-0.56}$ &  2.84$^{+0.39}_{-0.30}$ & 1670$\pm$130 \\ 
GC & \ldots & \ldots & 0.2 & \ldots &  0.04 & 1.59$\pm 0.16$ & 413$^{+128}_{-94}$ & 0.64 & 1.4 & 2.2 & 2.84 & 248$\pm 38$ \\
Field & \ldots& \ldots & 0.1 & \ldots & $<0.01$ & 1.89$\pm$0.11 & 448$^{+78}_{-69}$ & 0.03& 1.4 & 2.2 & 2.84 & 404$\pm 55$\\
\enddata
\tablecomments{The best-fitting parameters derived from fitting the XLF
of each galaxy, and the composite XLFs of all the sources (``composite'').
We also show fit results for the composite XLF of LMXBs in GCs (``GC'') 
and those in the field (``field'').
Listed are the number of sources used in fitting the XLF ($N_{src}$),
the fraction of the total galaxy light within the region in which the XLF
is computed ($f_{cover}$), the approximate minimum \lx\ of sources contributing
to the XLF ($L_{X,min}$), the approximate \lx\ of a source containing 20 photons
($L_{X,20}$), the null hypothesis probability (\probhnought), the differential 
logarithmic slope 
($\beta$), and the integrated luminosity of the X-ray sources in
the range \lx=$10^{37}$--$2\times 10^{39}$\ergps\ (see text). Fit results
are also shown for a broken power law model. For individual galaxies,
the slopes above and below the break ($\beta_2$ and $\beta_1$, respectively)
and the break luminosity, $L_{break}$, were fixed at values obtained
for the composite XLF. For each galaxy, we 
list in the $\beta_2^{free}$ column the value of this parameter if allowed
to fit freely, when it could be constrained. Error bars are quoted at the
90\% confidence limit. Where no error-bars are quoted on fit parameters,
they were fixed at the given value.}
\end{deluxetable*}
\subsection{Source detection incompleteness} \label{sect_incompleteness}
In order to measure the point-source XLF it is extremely important to 
take into account the effects of source detection incompleteness
and the Eddington bias, where the flux measurement errors distort
the XLF shape \citep[\eg][]{kim04b}. 
\citet{kim03a} presented a possible strategy to correct the data 
for these effects. In order to maintain the statistical
integrity of the data, we preferred to apply a correction to the model,
and so adopted a modified strategy, which incorporates some aspects of 
the algorithm of \citet{wang04a}.

For each galaxy, we performed a set of Monte-Carlo simulations,
initially using 20 flux bins (corresponding approximately to 
equally-spaced bins from 1--20 detected counts), which entailed
adding simulated sources to the image of each galaxy and then assessing
the performance of the detection algorithm at finding and characterizing
them. For many of the galaxies, we found that almost all of the sources
expected to contain $\sim$20 photons are detected, and so we assumed
that any brighter sources will always be detected. In those galaxies
for which this was not the case, we added additional simulations,
logarithmically spaced in flux space, until $\sim$95\%\ completeness
was achieved in any bin.
In each flux bin, we performed 20 such Monte-Carlo simulations,
in each of which we added 20 sources; adding them 20 at a time does not 
significantly increase source confusion since most galaxies had many
more than 20 LMXBs in the total field of view.
We assumed that the point-sources were distributed approximately as the 
optical light (\S~\ref{sect_spatdist}).
For each simulated source, to determine the number 
of photons to simulate
we computed a count-to-flux conversion by creating local ancillary spectral
response files and an on-axis redistribution matrix and
folding our preferred powerlaw spectral model through them.
The point-source images were derived at each point by bilinear interpolation 
of the \caldb\ PSF images for a 1~keV point-source at various focal-plane
positions, and finally Poisson noise was added.

In any arbitrary region of the \chandra\ field of view, for
each flux bin this yielded the fraction of simulated sources which
were detected and the {\em measured} count-rates (hence fluxes) 
of these sources. For brighter
sources than those simulated, we assumed 100\% completeness and that the 
distribution of measured fluxes is given by the Poisson statistics.
Since the XLF comprises a histogram of 
the numbers of sources detected in a series of narrow flux bins, this 
information is sufficient to construct a ``response matrix'' through
which any model to fit the XLF can be folded, analogous to  X-ray 
spectral-fitting
\citep[\eg][]{davis01a} to take account of both incompleteness and 
the Eddington bias. We note that this method also applies the 
same correction to the interlopers, which is formally incorrect since their
spatial distribution is much flatter than the
optical light.  However, there were few interlopers in the
regions of interest so this should not significantly affect our results. 

\subsection{XLF fitting} \label{sect_xlf}
\begin{figure*}
\centering
\includegraphics[scale=0.8]{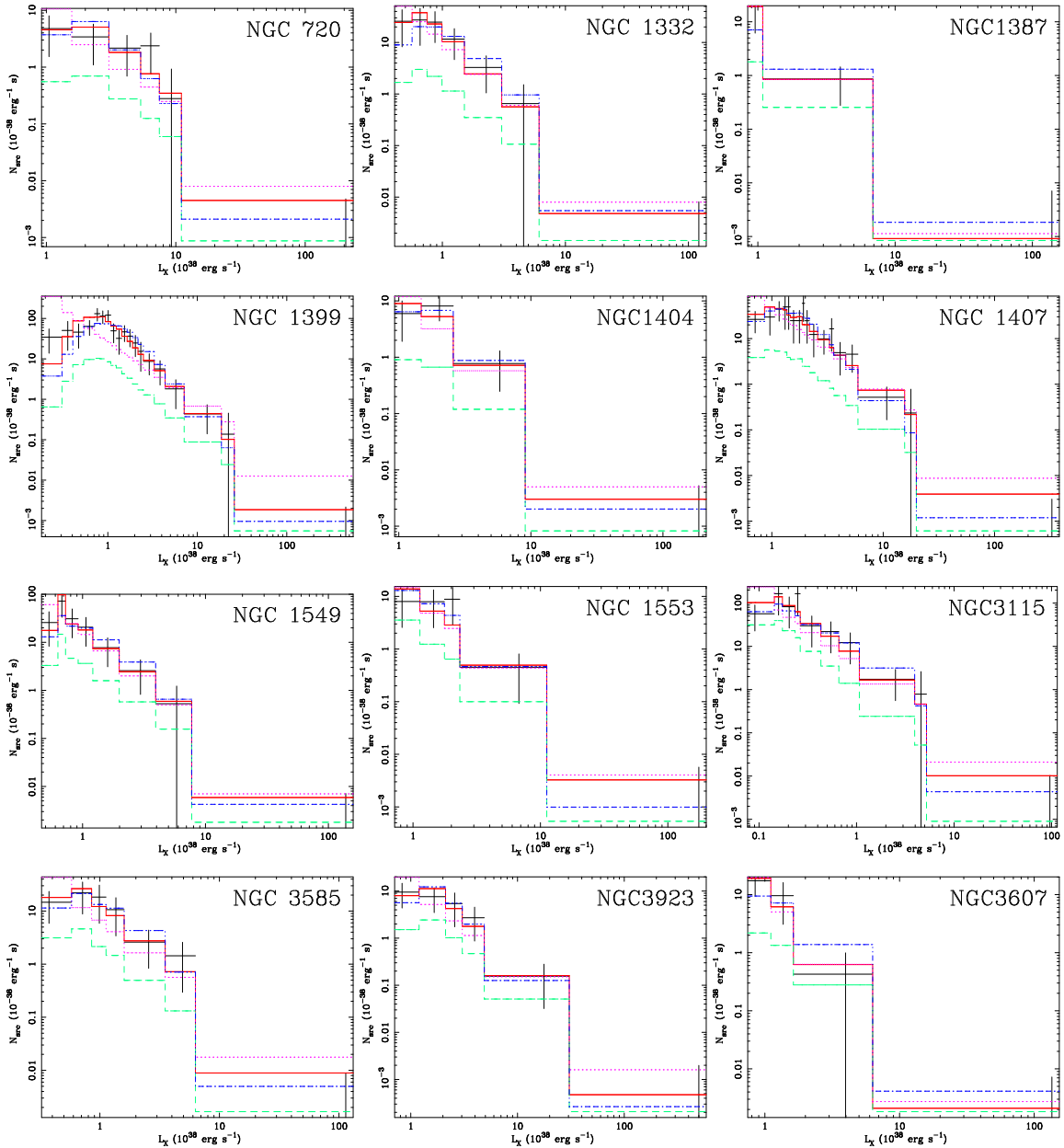}
\caption{Individual differential XLFs for half of the galaxies (continued
in Fig~\ref{fig_xlf2}), shown with the
best-fitting power law model (solid line; red), corrected for source 
detection incompleteness and the Eddington bias. The expected contribution
from background AGN is shown as a dashed line (blue-green) as is the best-fit
model to the composite XLF of all the galaxies (dash-dot-dot-dot; blue).
For comparison, the best-fit simple power law model, without incompleteness
correction, is also shown (dotted line; magenta).} \label{fig_xlf}
\end{figure*}
\begin{figure*}
\centering
\includegraphics[scale=0.8]{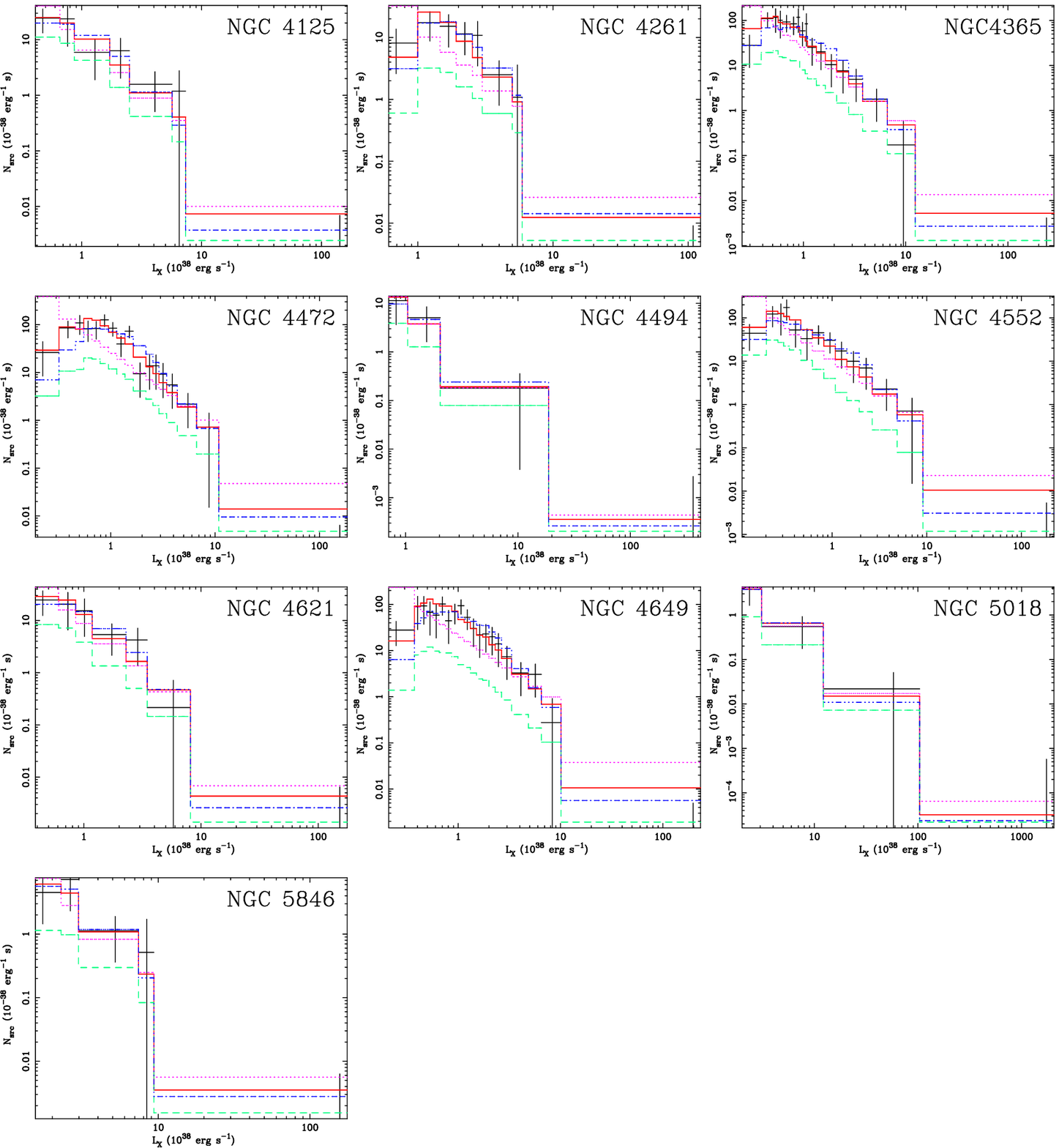}
\caption{Same as Fig~\ref{fig_xlf} for the remaining galaxies.} \label{fig_xlf2}
\end{figure*}
To measure the XLFs of the galaxies, we computed histograms
of the measured numbers of sources as a function of flux.
We considered only point-sources within \dtwentyfive\ and excluded the 
innermost 20\arcsec (where source detection may be uncertain;
\S~\ref{sect_spatdist}).
The data were regrouped to ensure at least 5 sources per bin, and the 
differential XLFs fitted using dedicated software which folded the 
model through the response described in \S~\ref{sect_incompleteness}.
To perform the fit we adopted the Cash-C statistic 
and parameter space was searched for a global minimum. To assess the
goodness-of-fit, we adopted a Monte-Carlo strategy. Adopting the 
best-fitting model, we simulated 100 fake data-sets which were subsequently
fitted. We adopted the  fraction of simulations which yielded a best-fitting 
Cash statistic value higher (\ie\ worse) than obtained for the real data as the 
null hypothesis (the model describes the data) probability (\probhnought).
To account for interlopers, we adopted the hard-band XLF relation for sources 
in the Chandra Deep Field-- South \citep{tozzi01b}, correcting to the 
spectral-fitting band used in the present work, 
and the area of sky under scrutiny. Since the slope of the 
interloper XLF at high luminosities is similar to that of 
a typical early-type galaxy \citep{kim04b}, it was not possible to fit
the background normalization freely and so it was fixed. To investigate
the sensitivity of our results to this assumption, we have experimented
with the effect of varying the normalization by $\pm$50\%. We found
that this typically caused changes in the best-fitting parameters
which were smaller than the statistical errors.

We found that we were generally able to fit the XLF of each galaxy
adequately with a simple powerlaw model of the form:
\begin{equation}
\frac{d N}{d L} \propto L^{-\beta}
\end{equation}
The best-fitting results are shown in Table~\ref{tab_xlf}, and the 
XLFs are shown in Figs~\ref{fig_xlf}--\ref{fig_xlf2}. In Table~\ref{tab_xlf} we list
the total luminosity in the range $10^{37}$--$2\times 10^{39}$\ergps\
determined from our fit, having been corrected upwards by the fraction
of optical light which falls outside the region in which the XLF 
was computed. This correction allowed us approximately to take account
of the fact that not all the LMXBs are expected to lie in this region,
since the spatial distribution of the LMXBs is generally close to
that of the optical light.  For NGC\thin 5845 there were
insufficient sources within the fitting region, and so we omit it
from this table.
For the 7 galaxies in our sample which overlapped that of \citet{kim04b},
we found good agreement, within the error-bars, between the measured
values of $\beta$.
\citet{xu05a} obtained 
an incompleteness-corrected fit for NGC\thin 4552, but found
a slightly steeper $\beta$ ($\sim$2.2) than we obtained. The reason
for the discrepancy is unclear, although these authors fitted the 
cumulative luminosity function which is difficult to interpret 
due to correlations between adjacent data-points. Given that a 
single powerlaw model was typically an adequate fit, we found that
we were generally unable to constrain the parameters of a broken
powerlaw fit to the data. 

\begin{figure}
\centering
\includegraphics[scale=0.35]{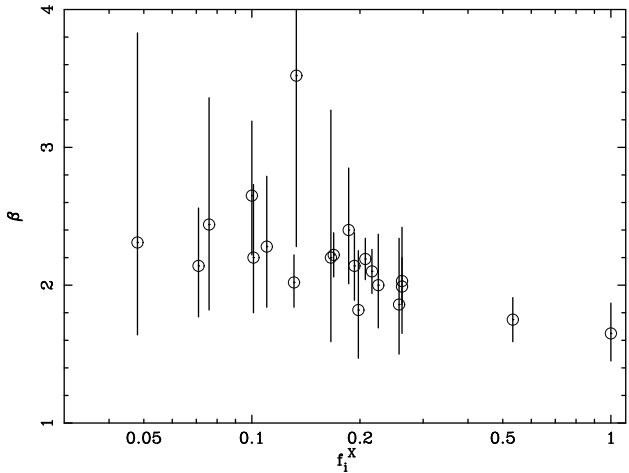}
\caption{Plot of XLF slope, $\beta$ as a function of source 
detection completeness, $f_i^X$. Note the correlation.} 
\label{fig_beta_inc}
\end{figure}
There was remarkably little variation in $\beta$,
indicating striking uniformity in the XLF shape of LMXBs in 
early-type galaxies. However, there is evidence of a small, but
statistically-significant variation in slope.
In Fig~\ref{fig_beta_inc} we show how $\beta$ varies as a function
of the X-ray source detection completeness, $f_i^X$, defined as the 
fraction of actual sources with \lx$=10^{37}$--$2\times 10^{39}$ \ergps\
which are detected. This
is obtained by integrating the completeness-corrected and the 
uncorrected XLFs over the appropriate flux range. To prevent spurious
correlations, we fixed $\beta=2.0$ for this calculation. 
The data show a clear anti-correlation; using 
Spearman's rank-order
correlation test, we found a probability of $\sim 7\times 10^{-5}$ that 
the data are uncorrelated. This correlation  indicates that the
XLF, on average, must flatten at low \lx. If the ``true'' XLF is,
in fact, a broken powerlaw which flattens below the break then
the measured $\beta$ will be an ``average'' of the slopes
above and below it.
As the data become more incomplete there are fewer
sources below the break to weight the fit, so $\beta$ asymptotes to
the high-\lx\ slope. Since $\beta$, where the data are very complete,
appears to asymptote to a value significantly flatter than 2.0, we
conclude that $\beta$ at low luminosities must, similarly, be 
significantly less than 2.0. 
We note that, if we exclude the two most complete data-points from
this comparison, we find that the probability of no correlation is still
$\sim 2\times10^{-3}$, indicating that this result is not solely driven
by these data.

\subsection{Composite XLF} \label{sect_composite_xlf}
\begin{figure}
\centering
\includegraphics[scale=0.35]{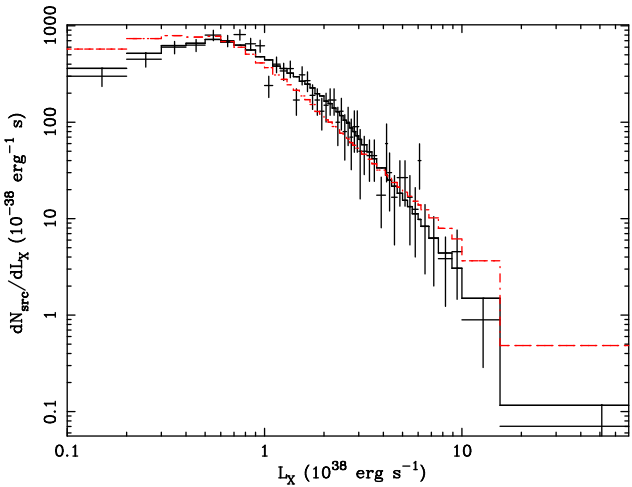}
\caption{Composite XLF of all LMXBs, fitted with a simple power law
(dashed line) and a broken power law (solid line). The models
have been corrected for source detection incompleteness and
the Eddington bias.} \label{fig_composite_xlf}
\end{figure}
Given the trend for systematically flatter XLFs with increasing
source detection completeness, we next 
experimented with adding the XLFs in order to
improve S/N. We combined all detected sources for all of the galaxies
into a single source list, and used that to generate the luminosity
function data-points. To account for source
detection incompleteness and the Eddington bias, we performed a
weighted addition of the individual responses generated for each galaxy,
using the number of observed sources as the weighting factors.
We grouped the data identically to \S~\ref{sect_xlf}, and similarly adopted 
the 
Cash-C fit-statistic.

We first fitted a single power law to the data, including a component
to account for the expected background sources (estimated by adding
the contributions anticipated for each galaxy).
We found this model to be a very poor fit to the data
(Fig~\ref{fig_composite_xlf}). The best-fitting $\beta$ (1.7) was
much lower than we obtained for each galaxy individually. This is 
easily understood since this shallow slope is largely driven by 
the apparent paucity of sources at low 
\lx\ (\ltsim 6$\times 10^{37}$\ergps), consistent with the XLF flattening
at low \lx.
We next tried to fit the data with
the truncated powerlaw model of \citet{sivakoff03}, which gave a break
at 6.7$\times 10^{38}$\ergps\ but failed to reproduce very well the shape of the
XLF at low luminosities (\probhnought$<0.1$\%).
Finally, we fitted a broken powerlaw to the composite XLF. 
This gave a significantly better fit to the data; although the formal 
\probhnought\ is only 3\%, the deviations from a good fit are primarily
due to individual discrepant data-points which are unlikely to bias the fit.
The best-fitting parameters are shown in Table~\ref{tab_xlf},
and the best-fit model is shown in Fig~\ref{fig_composite_xlf}.

We caution that, at the lowest luminosities the composite XLF
is largely dominated by two galaxies, NGC\thin 3115 and NGC\thin 4552 
and there exists a possibility that 
the XLFs of these two systems are not representative
of the whole population of early-type galaxies. If that is the case, it may
bias the shape we find for the composite low-\lx\ XLF. With our current data
it is not possible entirely to rule out such a possibility; instead it must
be tested using very deep \chandra\ observations of other galaxies. In the
present work we take this fit at face value, but are careful to
investigate the sensitivity of any derived quantities to the low-\lx\ slope.

To ensure that the broken power law fit did not arise as an 
artifact of our analysis, we experimented with simulating a simple power 
law XLF appropriate for each galaxy, with $\beta=2.0$. Combining these
identically to the real data, we found that a simple power law model
gave a good fit to the data. 
Conversely, if we simulated simple power law XLFs but with $\beta$ fixed
to the best-fit value for each galaxy, the composite XLF was very poorly
fitted by a simple power law, and instead a broken power law was required.
However the implied break luminosity 
($\sim 1.3 \times 10^{39}$\ergps) was considerably higher than that of
the composite XLF and the best-fit model for the real data was
an extremely poor fit.

Since the response matrices we used to correct for the effects of 
incompleteness were generated by adding a {\em finite} number of 
artificial sources to each image, this introduces a statistical 
error into the matrix, which we have not explicitly taken into 
account in our analysis. In order to assess whether the magnitude of 
this effect can lead to significant errors in our measured XLF shape,
we performed 100 Monte-Carlo simulations, in which all the ``response matrices''
were re-created, then combined and used to fit the existing data.
In the interests of speed, we did not 
re-create the matrices from first principles (as described in \S~\ref{sect_incompleteness}),
but instead, for each luminosity bin of artificial sources which were
added to the images, we drew 100 random numbers, which were distributed
in the same manner as the {\em observed} luminosities in this bin 
(including the fraction of non-detections). Treating these random values
as though they were the recovered luminosities of artificial point-sources
added to the image, we constructed a new ``fake'' response matrix.
Although this procedure assumes that the ``real'' matrix is formally 
correct, the differences in the faked matrices should accurately indicate
the magnitude of the statistical errors in that matrix.
We re-fitted the composite XLF with appropriately added ``faked'' responses
for each Monte-Carlo simulation, and assessed the 1-$\sigma$ scatter in the
best-fitting parameters, indicating the level of this source of error.
We found errors of $\pm 0.02$, 0.06$\times 10^{38}$\ergps\ and 0.02 for $\beta_1$, $L_{break}$
and $\beta_2$, respectively, which are considerably smaller than the 
statistical errors reported in Table~\ref{tab_xlf}.

If the XLFs of all the galaxies in the sample are essentially the same,
and it is simply the lack of counting statistics which have enabled
us to fit single power law models to each dataset individually 
(\S~\ref{sect_xlf}),
we would expect that the best-fit composite XLF model 
should also provide a good fit for each galaxy. We list in 
Table~\ref{tab_xlf}
the quality of fit for this model when applied to each galaxy in turn,
and the best-fit models are shown in Figs~\ref{fig_xlf}--\ref{fig_xlf2}. 
On a case-by-case basis, we find some systems which marginally seem 
inconsistent with having a break at $\sim 2\times 10^{38}$\ergps;
for example NGC\thin 1399.
In contrast, the broken powerlaw model seems marginally preferred
in some other systems, such as NGC\thin 4552 and NGC\thin 1404.
However, in none of these cases is the data of sufficient
quality to distinguish strongly between the two models.
On average the powerlaw and broken powerlaw fits seem to fit the 
data comparably well. Nonetheless, if either model is 'correct', the
\probhnought\ values listed for that model in Table~\ref{tab_xlf} should
be distributed uniformly between 0 and 1. We therefore tested this 
hypothesis with a K-S test. For the simple powerlaw model, the probability
that \probhnought\ are distributed uniformly is 1\%. In contrast, for
the broken powerlaw model, this probability is a more acceptable 65\%.

Throughout the rest of the paper, unless otherwise stated, 
we assume that the XLF of all galaxies is the same as our best-fitting 
model to the composite XLF, and extends down to 10$^{37}$ \ergps. In practice,
this means that all the  'completeness-corrected' results presented in
\S~\ref{sect_gcs} refer only to LMXBs with \lx $> 10^{37}$ \ergps.
Since the very low \lx\ XLF is dominated by only two galaxies, which could
bias its shape if they are not representative, we also assess how our 
conclusions would be affected by adopting an XLF shape which is much 
steeper in that luminosity regime (specifically an unbroken powerlaw model
with $\beta=2.0$).

\section{Globular cluster associations} \label{sect_gcs}
\renewcommand{\tabcolsep}{1mm}
\begin{deluxetable*}{llrlllrllrll}
\tablecaption{LMXB-GC properties \label{tab_gcs}}
\tabletypesize{\small}
\tablehead{
\colhead{Name }&\colhead{${L_V^{FOV}}$}&\multicolumn{2}{c}{$N_{GC}$}&\colhead{${L^{TOT}_{GC}}$}&\colhead{$f^{GC}_i$} &\colhead{$N_{X,obs}$}&\colhead{$N_X$}&\colhead{$f_i^X$}&\multicolumn{2}{c}{$N_{GC,X}$}&\colhead{${N_{false}}$}\\
\colhead{ }&\colhead{}&\colhead{red}&\colhead{blue}&\colhead{}&\colhead{ }&\colhead{ } & \colhead{} & \colhead{} & \colhead{red} & \colhead{blue} & \colhead{}\\
\colhead{ }&\colhead{($10^{10}$\lsun)}&\colhead{}&\colhead{ }&\colhead{($10^7$\lsun)}&\colhead{ }&\colhead{ } & \colhead{} & \colhead{} & \colhead{} & \colhead{} & \colhead{}}
\startdata
NGC1332, &$2.0\pm 0.5$ &$65$ &$133$ &$8.3\pm 1.2$ &$0.60$ &$27$ &$86.\pm 19.$ &$0.29$ &$3$ &$3$ &$0.70$ \\
NGC1387, &$1.2\pm 0.2$ &\multicolumn{2}{c}{28} &$0.79\pm 0.51$ &$0.22$ &$10$ &$26.\pm 10.0$ &$0.32$ & \multicolumn{2}{c}{0} &$0.0$ \\
NGC1399, &$2.0\pm 1.1$ &$162$ &$329$ &$12.\pm 1.$ &$0.81$ &$42$ &$160\pm 30$ &$0.27$ &$7$ &$13$ &$1.5$ \\
NGC1404, &$1.7\pm 1.1$ &$47$ &$116$ &$5.3\pm 0.9$ &$0.74$ &$11$ &$58.\pm 17.$ &$0.19$ &$0$ &$5$ &$0.20$ \\
NGC1553, &$2.7\pm 0.3$ &$21$ &$50$ &$2.1\pm 0.5$ &$0.76$ &$13$ &$23.\pm 8.$ &$0.48$ &$0$ &$2$ &$0.20$ \\
NGC3115, &$1.08\pm 0.08$ &$39$ &$93$ &$2.3\pm 0.4$ &$0.92$ &$33$ &$31.\pm 6.$ &$0.91$ &$7$ &$2$ &$0.60$ \\
NGC3585, &$3.0\pm 0.5$ &$35$ &$67$ &$2.4\pm 0.4$ &$0.83$ &$21$ &$40.\pm 10.$ &$0.45$ &$3$ &$0$ &$0.30$ \\
NGC3607, &$2.4\pm 0.5$ &$21$ &$85$ &$4.9\pm 0.9$ &$0.62$ &$19$ &$58.\pm 15.$ &$0.30$ &$1$ &$3$ &$0.40$ \\
NGC4125, &$3.2\pm 0.5$ &$77$ &$118$ &$2.6\pm 0.4$ &$0.74$ &$23$ &$38.\pm 9.$ &$0.50$ &$1$ &$1$ &$0.80$ \\
NGC4261, &$3.4\pm 2.0$ &$70$ &$170$ &$11.\pm 1.$ &$0.56$ &$17$ &$78.\pm 18.$ &$0.20$ &$1$ &$3$ &$0.20$ \\
NGC4365, &$2.3\pm 0.4$ &$90$ &$200$ &$8.2\pm 0.9$ &$0.81$ &$42$ &$71.\pm 12.$ &$0.53$ &$6$ &$9$ &$1.6$ \\
NGC4472, &$3.9\pm 0.3$ &$183$ &$214$ &$7.1\pm 0.7$ &$0.83$ &$37$ &$82.\pm 14.$ &$0.43$ &$7$ &$6$ &$0.70$ \\
NGC4494, &$1.5\pm 0.3$ &$42$ &$133$ &$3.0\pm 0.4$ &$0.86$ &$8$ &$19.\pm 7.$ &$0.39$ &$4$ &$0$ &$1.6$ \\
NGC4552, &$1.3\pm 0.2$ &$77$ &$162$ &$4.2\pm 0.5$ &$0.85$ &$36$ &$56.\pm 10.$ &$0.56$ &$5$ &$6$ &$1.3$ \\
NGC4621, &$1.7\pm 0.3$ &$57$ &$114$ &$3.9\pm 0.5$ &$0.86$ &$27$ &$44.\pm 9.$ &$0.53$ &$2$ &$8$ &$0.80$ \\
NGC4649, &$2.9\pm 0.3$ &$130$ &$192$ &$7.2\pm 0.7$ &$0.87$ &$35$ &$91.\pm 15.$ &$0.38$ &$13$ &$4$ &$1.2$ \\
NGC5018, &$9.7\pm 2.5$ &$10$ &$80$ &$7.8\pm 1.8$ &$0.47$ &$8$ &$49.\pm 28.$ &$0.095$ &$0$ &$1$ &$0.20$ \\
NGC5845, &$1.9\pm 0.7$ &$18$ &$39$ &$1.4\pm 0.8$ &$0.47$ &$2$ &$0.32\pm 3.9$ &$0.20$ &$0$ &$0$ &$0.0$ \\
NGC5846, &$2.6\pm 0.7$ &$112$ &$175$ &$5.3\pm 0.6$ &$0.76$ &$8$ &$37.\pm 14.$ &$0.19$ &$2$ &$1$ &$0.40$ \\
\enddata
\tablecomments{Details of the GC populations within the WFPC2 field of view.
We show the total V-band luminosity in the field of view ($L_V^{TOT}$),
the number of GCs ($N_{GC}$), divided into red and blue GCs (except for 
NGC\thin 1387, for which we do not have colour information; \gcpaper),
the total GC luminosity ($L^{TOT}_{GC}$), corrected for 
incompleteness, the GC source detection 
completeness, $f_i^{GC}$, which is the fraction of the total GC luminosity
detected.
We also list the number of LMXBs observed in the WFPC2 field of view 
($N_{X,obs}$), the (incompleteness-corrected) number of such sources in
the WFPC2 field of view ($N_{X}$), the fraction of the LMXBs which are detected
($f_i^X$; this is derived from our fits, not strictly from taking $N_{X,obs}/N_X$,
which is more affected by statistical noise), the number of LMXBs with 
corresponding GC counterparts ($N_{GC,X}$) and the number of probable false
GC-LMXB matches ($N_{false}$).
}
\end{deluxetable*}
\renewcommand{\tabcolsep}{1mm}
\begin{deluxetable*}{llrlrlrl}
\tabletypesize{\tiny}
\tablecaption{Derived LMXB-GC properties\label{tab_gcs2}}
\tablehead{
\colhead{Galaxy}&\colhead{\Sl} & \multicolumn{2}{c}{\Sxn}&\multicolumn{2}{c}{$p_{GC,X}$}&\multicolumn{2}{c}{LMXB GC fraction} \\
\colhead{ }&\colhead{ }&\colhead{corrected} & \colhead{raw} &\colhead{corrected}&\colhead{raw}&\colhead{corrected}&\colhead{raw} 
}
\startdata
NGC1332 &$0.42\pm 0.19$ &$0.38\pm 0.08$ &$0.11\pm 0.03$ &$0.018^{+0.012}_{-0.008}$ &$0.027\pm 0.018$ &$0.35^{+0.24}_{-0.16}$ &$0.20^{+0.13}_{-0.09}$ \\
NGC1387 &$0.065\pm 0.045$ &$0.18\pm 0.07$ &$0.071\pm 0.030$ &$<$$0.18$ &$<$0.05 &$<$$1.0$ &$<$$0.19$ \\
NGC1399 &$0.59\pm 0.53$ &$0.65\pm 0.11$ &$0.17\pm 0.03$ &$0.039^{+0.012}_{-0.009}$ &$0.038^{+0.011}_{-0.009}$ &$0.55\pm 0.16$ &$0.43\pm 0.13$ \\
NGC1404 &$0.32\pm 0.35$ &$0.30\pm 0.09$ &$0.061\pm 0.023$ &$0.037^{+0.026}_{-0.017}$ &$0.029^{+0.021}_{-0.013}$ &$0.58^{+0.41}_{-0.27}$ &$0.40^{+0.28}_{-0.18}$ \\
NGC1553 &$0.077\pm 0.024$ &$0.071\pm 0.024$ &$0.040\pm 0.015$ &$0.013^{+0.019}_{-0.009}$ &$0.025^{+0.037}_{-0.019}$ &$0.22^{+0.32}_{-0.16}$ &$0.14^{+0.20}_{-0.10}$ \\
NGC3115 &$0.21\pm 0.04$ &$0.24\pm 0.04$ &$0.27\pm 0.05$ &$0.026^{+0.013}_{-0.009}$ &$0.064^{+0.031}_{-0.022}$ &$0.33\pm 0.16$ &$0.25^{+0.12}_{-0.09}$ \\
NGC3585 &$0.081\pm 0.026$ &$0.12\pm 0.03$ &$0.060\pm 0.016$ &$0.015^{+0.016}_{-0.009}$ &$0.026^{+0.029}_{-0.016}$ &$0.18\pm 0.19$ &$0.13^{+0.14}_{-0.08}$ \\
NGC3607 &$0.20\pm 0.07$ &$0.20\pm 0.05$ &$0.066\pm 0.019$ &$0.026^{+0.023}_{-0.014}$ &$0.034^{+0.030}_{-0.018}$ &$0.33^{+0.29}_{-0.18}$ &$0.19\pm 0.17$ \\
NGC4125 &$0.082\pm 0.023$ &$0.10\pm 0.02$ &$0.062\pm 0.016$ &$<$$0.015$ &$<$$0.020$ &$<$$0.27$ &$<$$0.16$ \\
NGC4261 &$0.34\pm 0.34$ &$0.20\pm 0.05$ &$0.042\pm 0.013$ &$0.015^{+0.012}_{-0.007}$ &$0.016^{+0.013}_{-0.008}$ &$0.42^{+0.35}_{-0.22}$ &$0.22\pm 0.19$ \\
NGC4365 &$0.36\pm 0.11$ &$0.27\pm 0.05$ &$0.16\pm 0.03$ &$0.020^{+0.007}_{-0.006}$ &$0.046\pm 0.017$ &$0.44\pm 0.16$ &$0.31^{+0.12}_{-0.09}$ \\
NGC4472 &$0.18\pm 0.03$ &$0.18\pm 0.03$ &$0.082\pm 0.016$ &$0.019^{+0.007}_{-0.006}$ &$0.031^{+0.012}_{-0.009}$ &$0.42\pm 0.16$ &$0.32^{+0.12}_{-0.09}$ \\
NGC4494 &$0.20\pm 0.07$ &$0.11\pm 0.04$ &$0.037\pm 0.023$ &$0.014\pm 0.019$ &$0.014\pm 0.018$ &$0.37^{+0.50}_{-0.30}$ &$0.30^{+0.40}_{-0.24}$ \\
NGC4552 &$0.32\pm 0.10$ &$0.37\pm 0.06$ &$0.23\pm 0.05$ &$0.024^{+0.011}_{-0.008}$ &$0.040\pm 0.018$ &$0.36\pm 0.17$ &$0.27^{+0.12}_{-0.09}$ \\
NGC4621 &$0.23\pm 0.07$ &$0.22\pm 0.05$ &$0.14\pm 0.03$ &$0.026^{+0.012}_{-0.009}$ &$0.053^{+0.025}_{-0.018}$ &$0.46^{+0.22}_{-0.16}$ &$0.33\pm 0.15$ \\
NGC4649 &$0.25\pm 0.04$ &$0.27\pm 0.04$ &$0.098\pm 0.020$ &$0.031^{+0.010}_{-0.008}$ &$0.049\pm 0.016$ &$0.53\pm 0.17$ &$0.45\pm 0.15$ \\
NGC5018 &$0.080\pm 0.038$ &$0.043\pm 0.025$ &$(6.9  \pm 3.4) \times 10^{-3}$ &$<$$0.073$ &$<$$0.035$ &$<$$1.4$ &$<$$0.39$ \\
NGC5845 &$0.077\pm 0.062$ &$(1.5  \pm 18.) \times 10^{-3}$ &$(9.2  \pm 12.) \times 10^{-3}$ &$<$$0.060$ &$<$$0.033$ &$<$$62.$ &$<$$0.93$ \\
NGC5846 &$0.21\pm 0.09$ &$0.12\pm 0.05$ &$0.025\pm 0.013$ &$0.016\pm 0.018$ &$(9.0^{+10.}_{-5.8}) \times 10^{-3}$ &$0.50^{+0.56}_{-0.32}$ &$0.32^{+0.37}_{-0.21}$ \\
\enddata
\tablecomments{Derived properties for the LMXB and GC populations of the galaxies.
We list the GC specific luminosity, \Sl\ (\gcpaper) and the specific frequency
of LMXBs (\Sxn). We also include the probability that a given GC has an 
LMXB counterpart (\pgcx), and the fraction of LMXBs with a GC counterpart
(LMXB GC fraction). 
For \Sxn, \pgcx\ and the fraction of LMXBs in GCs, we list the value
corrected for incompleteness and GC colour effects (``corrected'') and a value derived 
directly from the data without any such correction. We stress that
\pgcx\ and the LMXB GC fraction are {\em not} directly fitted in our 
analysis, but we provide them here for convenience in interpreting 
Figs~\ref{fig_pgcx} and \ref{fig_pxgc}. The error-bars for \Sxn, \pgcx\
and LMXB GC fraction are 1-$\sigma$.
}
\end{deluxetable*}

\renewcommand{\tabcolsep}{0.5mm}
We discuss in \gcpaper\ our data-reduction and analysis of 
archival HST WFPC2 data for 19 of the galaxies in our sample.
In that paper, we provide detailed lists of all globular cluster 
candidates we detected, and relevant photometry and various
derived quantities, such as GC specific luminosity.
A number of the galaxies do not have WFPC2 data available, and so,
for consistency, we do not consider them here.
We consider only WFPC2 data for simplicity of our analysis, despite the 
availablity of superior ACS data for some of the galaxies. In practice,
useful ACS data are available only for less than half of the sample,
whereas most have WFPC2 data. Even with the relatively shallow WFPC2 data,
we were able to detect, in most objects, more than 75\% of the GC light,
which is sufficient for our present purposes.
We focus only on central-pointings with the WFPC2, although for
some of the galaxies there are multiple HST pointings. We adopt 
this procedure since the \chandra\ PSF rapidly degrades off-axis, making
matching GCs and LMXBs more challenging, and possibly introducing
a systematic bias between those objects with multiple pointings 
and those with only one.

Since the absolute astrometry of \chandra\ is not expected to be accurate at
sub-arcsecond precision, in order to assess possible matches between
LMXBs and GCs, we allowed a translation and a rotation transformation
between the GC and LMXB source lists, which were adjusted to maximize the
number of matches. For self-consistency, we assumed that any GC candidate
lying within 0.5\arcsec\ of a \chandra\ source centroid (comparable to the 
on-axis PSF) was associated with it. If, based on this criterion, 
more than one GC matches an LMXB, or if more than one LMXB matches a GC,
we match the pair with the closest centroids, although this only affects a 
small fraction of sources.
Where both the optical and X-ray images of the galaxy 
had clearly-defined centroids,
we constrained the transformation to ensure approximately the same matching
precision between them as between GCs and LMXBs. This 
typically had little effect in galaxies with a large number of LMXB-GC 
matches, but was an important constraint in other systems.
In order to assess the number of spurious matches obtained
by this procedure, we randomly generated fake GC source lists, scattering
them approximately as the optical light until the total number of 
objects in the WFPC2 FOV matched that which was observed. Using the 
procedures outlined above, we determined the number of these sources
apparently associated with an LMXB, which was typically less than $\sim$1.

\subsection{Composite spectra of GC sources}
We next investigated  potential spectral differences between LMXBs hosted
by red and by blue GCs.
Exactly analogous to the analysis in \S~\ref{sect_composite_spectra},
we obtained composite spectra for those LMXBs associated with red
GCs and those associated with blue GCs.
We defined any candidate GC with V-I$>$1.1 as red, and those
GCs with V-I$<$1.1 as blue\footnote{The colour at which between red and blue 
GC sub-populations are separated varies from galaxy to galaxy, but we suspect 
the GC properties of LMXB hosts depend on the metallicity rather than 
the sub-population to which they belong, hence we use a fixed V-I division
for all galaxies}. 
The spectra are shown in Fig~\ref{fig_spectra}
(along with the composite spectrum of all LMXBs in the galaxies).
Both spectra were well-fitted by the best-fit
models for the entire population (for the powerlaw models,
we obtained $\chi^2$/dof=109/137 and 100/99 for the red and blue GCs,
respectively. For the bremsstrahlung models $\chi^2$/dof was
111/137 and 102/99). There was no evidence of a systematically
harder spectrum or intrinsic absorption for the  blue-GC sources,
in conflict with the model of \citet{maccarone04a}.
Fitting simultaneously
the blue-GC and red-GC data with the same bremsstrahlung
model but allowing \nh\ to vary freely we obtained \nh${\rm <3\times 10^{20} cm^{-2}}$
for the blue-GC sources and $<10^{20} cm^{-2}$ for the red-GC sources.
The larger upper limit for the blue-GC case in part reflects the poorer
counting statistics but in both cases we can rule out intrinsic
absorption of the order of $10^{21} cm^{-2}$.
\subsection{GC and non-GC XLF} \label{sect_gc_xlf}
\begin{figure}
\centering
\includegraphics[scale=0.35]{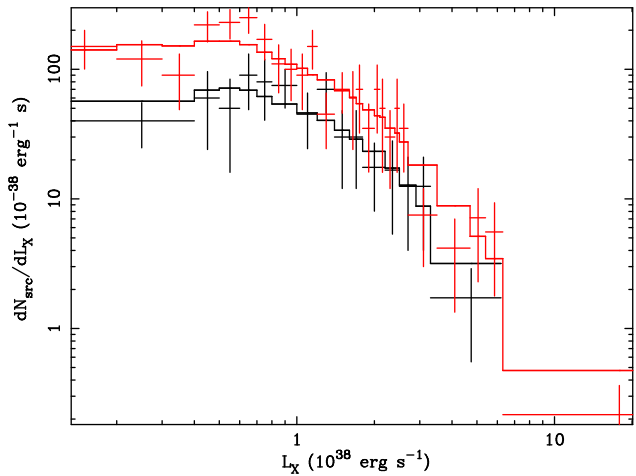}
\caption{Composite XLFs of GC LMXBs (black)
and field LMXBs (red), fitted with the best-fitting model
for the entire LMXB population.}
\end{figure}
We accumulated composite XLFs of those LMXBs associated
with GC and those which are in the WFPC2 field but not
identified with a GC, 
exactly as described in \S~\ref{sect_composite_xlf}.
In practice, we omitted any sources found in 
the very central regions of most of the galaxies
(as discussed in \gcpaper), since features in the optical images,
such as dust lanes, can give rise to false or ambiguous GC detections. In practice,
this involved excluding regions of radius between 2\arcsec\ and
8\arcsec. The XLFs
were remarkably similar in shape and both closely resembled
the composite XLF for the entire source population (which was
computed in a slightly different region). We therefore tried fitting
both data-sets with the preferred broken powerlaw model fitted in
\S~\ref{sect_composite_xlf}. Since the likelihood
of an interloper being spuriously matched to a GC candidate is 
low, we assumed that all of the expected interloper sources 
only contributed to the non-GC XLF. The broken powerlaw
model fitted  both data-sets adequately (Table~\ref{tab_xlf}),
indicating that the XLF of sources in GCs and those in the field
are essentially the same.

\subsection{Spatial distribution of GCs and LMXBs} \label{sect_gc_dist}
We showed in \S~\ref{sect_spatdist} that the spatial distribution of 
LMXBs closely follows that of red GCs (and also resembles the optical 
light outside the innermost $\sim0.5$\arcmin) in a subset of galaxies in
the sample. If some of the LMXBs observed in the field are born {\em in situ}, 
there may, nevertheless, be detectable differences between the
radial distributions of field and GC LMXBs.
Ideally we would like to address whether the spatial distributions
of LMXBs in GCs resembles that of the GC population, while for field LMXBs
it more closely resembles the optical light. However, given the limited
radial range over which we have measured the GC spatial distribution,
and the relatively small numbers of sources (both GCs and, in particular,
LMXBs) that are detected, this is impractical. Therefore, we instead
first tested, on a galaxy-by-galaxy basis and 
within the WFPC2 field of view, whether the 
radial distances  of the GC and field LMXBs from the galaxy centre
are consistent with being drawn 
from the same distribution. We assessed this with a Kolmogorov-Smirnov test
and only considered galaxies with at least
3 GC-LMXB matches. In all cases the hypothesis
that the two samples were the same could not be rejected at the 5\% 
significance level or better. 
Next, we applied the same test to the entire population of LMXBs
in the WFPC2 field, and similarly found a null hypothesis probability
of $\sim$64\%, indicating excellent agreement between the two
distributions.

\subsection{Probability that a GC contains an LMXB}
The probability that a GC is associated with an LMXB does not 
depend on the LMXB luminosity, as indicated by the similarities
in the XLF of the GC and non-GC X-ray sources.
Although in \gcpaper\ we compute the mass and metallicity of 
each GC, in order to investigate how the presence of an LMXB depends
upon the GC properties, we focus only on the luminosity and colour.
We adopt this approach for two reasons. Firstly, the metallicity errors 
are typically fairly large, and a non-negligible fraction of them
have metallicities which peg at the maximum or minimum values 
allowed in our conversion. Secondly, as we show in \gcpaper, the 
colour and luminosity of the GC populations are independent of 
each other, making this analysis simpler.
Since our optical data comprised, at most, 2 colours, we 
were not able to break the age-metallicity degeneracy and so
we assumed that all colour dependency reflects a metallicity 
effect \citep[c.f.][]{kundu03}. 
For consistency with past work, we make the conversion between
colour and metallicity by using the relation of \citet{kundu98a}, 
$log(Z_{Fe,GC}) = -5.89+4.72(V-I)$.
To investigate how the probability that a GC contains an active LMXB,
P(GC,X), depends upon the properties of the GC, we adopted the 
hypothesis:
\begin{equation}
P(GC,X) = p_{GC-X} \left(\frac{L_{GC}}{L_0}\right)^\alpha 
\left(\frac{Z_{GC}}{Z_0}\right)^\gamma \label{eqn_pgcx}
\end{equation}
where \lgc\ is the V-band luminosity of a GC, $Z_{GC}$
is its metallicity, $L_0$ is the luminosity of the turnover in the
globular cluster luminosity function (GCLF) for the entire 
GC population ($\sim 7.2\times 10^{4}$ \lsun), and $Z_0$ is the metallicity 
corresponding to the peak of the GC colour distribution ($\sim$0.08\zsun;
\gcpaper). The indices $\alpha$ and $\gamma$, and the normalizing
constant, $p_{GC-X}$, are to be determined by the data.

\subsubsection{Globular Cluster luminosity functions}
\begin{figure}
\centering
\includegraphics[scale=0.35]{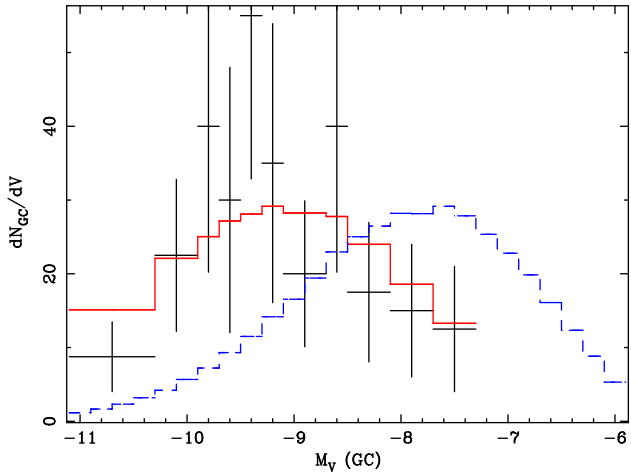}
\caption{Composite globular cluster luminosity functions
of all GCs associated with an LMXB.
The best-fit Gaussian model is shown (solid line), along with the
best-fit model to the GCLF of the entire population, rescaled
for clarity (dashed line). The models have been corrected for source detection 
incompleteness.} \label{fig_gclfs}
\end{figure}
Considering only those GC hosting an LMXB, we computed a 
globular cluster luminosity function (GCLF), to compare with
that of the entire population (\gcpaper).
We discuss in \gcpaper\ how the composite V-band GCLF is accumulated
and how we fitted the data to account for source detection
incompleteness.  We assumed that the detection incompleteness 
of GCs hosting LMXBs in any galaxy
depends upon luminosity in the same way as that of the whole
GC population. For simplicity, we omitted NGC\thin 1399 and NGC\thin 1404
from this calculation, since they did not have any V-band photometry.
We fitted the composite GCLF using a maximum-likelihood 
fitting algorithm and minimizing the Cash-C statistic exactly as described 
in \gcpaper. We found the GCLF of the X-ray luminous source population could
be fitted  by a single Gaussian distribution, \ie\
\begin{equation}
\frac{d n_{GC}}{d V} = \frac{N_{GC}}{\sqrt{2 \pi} \sigma} \exp
\left( -\frac{(V-V_T)^2}{2 \sigma^2}\right)
\end{equation}
where $n_{GC}$ is the number density of GCs, as a function of V-band 
luminosity (V), $N_{GC}$ is the total number of GCs and $V_T$ is the 
apparent V-band peak luminosity (``GCLF turnover'').
 Based on 100 Monte-Carlo
simulations, we estimated \probhnought=46\%.
We found $\sigma=1.0^{+0.4}_{-0.2}$, in  agreement with that of the entire 
population ($\sim$1.5),  but the absolute magnitude of the V-band turnover
was -9.0$^{+0.3}_{-0.2}$,
approximately 2.0 magnitudes brighter than the global GCLF,
consistent with previous observations \citep{kim05a}.
Fig~\ref{fig_gclfs} shows the GCLF,
along with the best-fitting Gaussian distribution and, for comparison
purposes, the best-fit Gaussian to the entire GC population,
suitably rescaled.

Since there is no correlation between
\lgc\ and $Z_{GC}$ (\gcpaper; \S~\ref{sect_gc_colours}), the dependence
of P(GC,X) on these quantities can be assessed independently by considering
separately the GCLF and the GC colour distributions. 
It follows from Eq~\ref{eqn_pgcx}  that the 
GCLF of X-ray hosting GCs will have the same shape as the entire
population, but be systematically shifted by 
$-0.92\alpha \sigma^2$ magnitudes. This implies that $\alpha=1.01\pm0.19$.

\subsubsection{Globular Cluster colour distributions} \label{sect_gc_colours}
\begin{figure}
\centering
\includegraphics[scale=0.35]{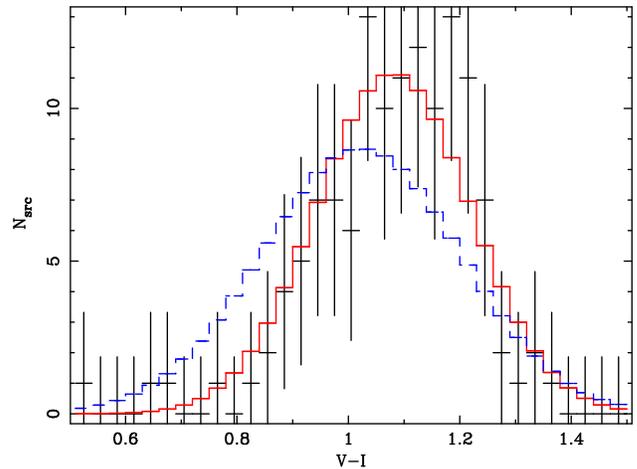}
\caption{Composite colour distribution histogram of all GCs 
associated with an LMXB. Also shown are the best-fit
Gaussian model (solid line; red) and the best-fit Gaussian model
for the entire GC population, suitably renormalized (dashed line; blue).} \label{fig_gc_colours}
\end{figure}
We constructed
histograms of the numbers of GC as a function of V-I, each bin having 
a width of 0.05 mag. The histogram could be adequately fitted
({\em via} a Cash-C minimization algorithm)
with  a single Gaussian model, \ie\
\begin{equation}
\frac{d n_{GC}}{d (V-I)} = \frac{N_{GC}}{\sqrt{2 \pi} \sigma_c} \exp
\left( -\frac{((V-I)-(V-I)_0)^2}{2 \sigma_c^2}\right)
\end{equation}
where $n_{GC}$ is the number density of GCs, as a function of V-I 
colour, $N_{GC}$ is the total number of GCs and $(V-I)_0$ is the 
mean of the colour distribution. We found $(V-I)_0 = 1.08\pm 0.02$~mag and
$\sigma_c=0.14\pm0.2$~mag.
This distribution is  significantly redder and narrower than 
the GC population as a whole (Fig~\ref{fig_gc_colours}; \gcpaper),
implying the ratio of the probability that a red GC hosts an
LMXB to the probability that a blue GC does is $\sim$2.
We note that fitting these data with a simple Gaussian model is 
overly-simplistic since early-type galaxies generally exhibit bimodal
GC colour distributions. However, the peaks of the metal-rich and metal-poor
subpopulations are not cleanly separated in V-I space, and so the average
colour distributions of the clusters in a galaxy can be parameterized
by a simple Gaussian, which is sufficiently accurate for our present purposes.

The difference in $\sigma$ between the population of GCs as a whole
and the subpopulation hosting LMXBs is easily understood in terms
of the statistical errors on the photometry, since luminous GCs
preferentially host LMXBs.
If we considered only those GCs with absolute V-band magnitude $<$-9.0,
we found $\sigma=0.14$, in good agreement with the colour distributions
of the LMXB hosts. The fainter GCs will have increasing statistical
errors on their photometry, which will increase $\sigma$. 
Assuming no correlation between the colour and magnitude of a GC
(\gcpaper), it is trivial to show  from Eq~\ref{eqn_pgcx} and our adopted
colour-metallicity relation that the colour distribution of 
GCs hosting LMXBs should have the same shape as that of the 
entire population, but be shifted by $10.87\sigma^2 \gamma$
magnitudes, implying $\gamma=0.33\pm0.11$.

\subsubsection{Probability that a GC hosts an LMXB} \label{sect_pgcx}
\begin{figure}
\centering
\includegraphics[scale=0.34]{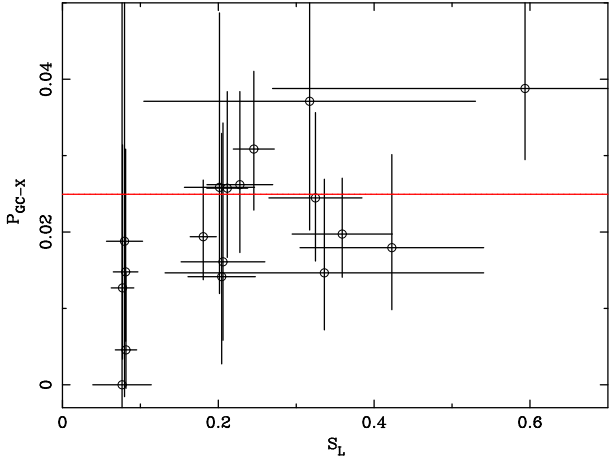}
\caption{Incompleteness, colour and luminosity 
corrected fraction of GCs which contain an 
LMXB (see Eq~\ref{eqn_pgcx}), shown as a function of 
\Sl. We also show the best-fitting model.} 
\label{fig_pgcx}
\end{figure}
Given the shapes of the GCLFs and GC colour distributions
for our sample (for our purposes the latter can be approximated
as a Gaussian with a peak shifted by $\Delta_c$ magnitudes from
that of the entire sample, and with width $\sigma_c$), 
it follows from  Eq~\ref{eqn_pgcx}
that 
\begin{eqnarray}
N_{GC,X}-N_{false} = p_{GC-X} \frac{L^{TOT}_{GC}}{L_0} f^{X}_i f^{GC}_{i} 
\nonumber \\ \times \exp (59.0 \gamma^2 \sigma_c^2 +
10.9 \gamma \Delta_c)\label{eqn_nmatch}
\end{eqnarray}
where $N_{GC,X}$ is the number of GC-LMXB matches, $N_{false}$ 
the number of expected false matches, $L^{TOT}_{GC}$ the total
luminosity of GCs, $L_0$ the peak luminosity of the composite GCLF
of the entire sample, $f_i^X$ is  the fraction of LMXBs which are
detected and $f_i^{GC}$ is the fraction of the GC luminosity which 
is detected. For simplicity we here assumed that $\alpha=1$, with
which our data are consistent.
For NGC\thin 1387 we had no colour information for the GCs,
and so we adopted $\Delta_c=0.0$ and $\sigma_c=0.14$.

Using the information in Tables~\ref{tab_gcs} and \ref{tab_gcs2}, for any
given value of $p_{GC-X}$ Eq~\ref{eqn_nmatch} gives the number
of expected GC-LMXB matches. Matching this model to the observed
numbers of sources {\em via} the Cash-C statistic, and adopting
a maximum-likelihood fitting procedure similar to that outlined in
\S~\ref{sect_xlf} we were able to place constraints on $p_{GC-X}$.
We obtained
a good fit to the data (\probhnought=63\%) with \pgcx=$0.025\pm0.004$.
Integrating Eq~\ref{eqn_pgcx} over the mean colour and luminosity
distributions for the GC populations, this corresponds to a probability
of 6.5\% that a randomly chosen GC contains an LMXB with \lx$>10^{37}$ \ergps.
We show in Fig~\ref{fig_pgcx} \pgcx\ computed directly from
Eqn~\ref{eqn_nmatch} for each dataset
and the best-fit model value, as a function of GC specific luminosity, \Sl.
For comparison, if we replaced $f_i^X$ with the equivalent value 
appropriate for a powerlaw XLF, assuming $\beta=2.0$,
we also obtained an acceptable fit (\probhnought=29\%) with
\pgcx=$0.054\pm0.009$ (implying~14\% of GC contain an LMXB with 
\lx$> 10^{37}$ \ergps).

\subsection{Are any LMXBs formed in the field?} \label{sect_field_lmxb}
\begin{figure}
\centering
\includegraphics[scale=0.35]{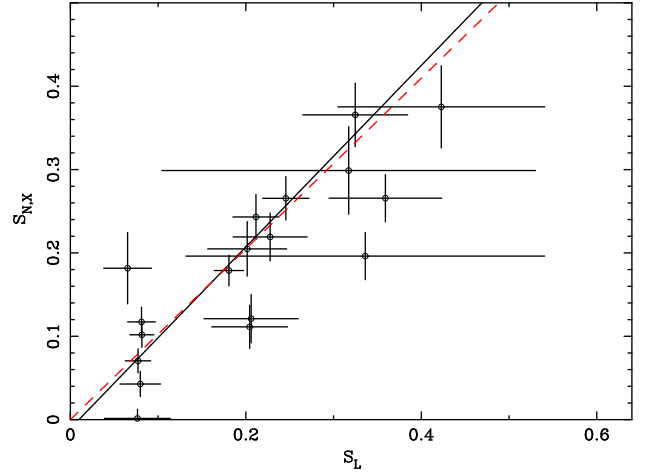}
\caption{V-band specific frequency of LMXBs (\Sxn) {\em versus}
specific luminosity of GCs (\Sl) for the WFPC2 FOV. The best-fit
straight-line relation is shown (solid line), along with the 
best-fit line if the intercept is fixed at 0 (dashed line; red)} 
\label{fig_nx_ngc}
\end{figure}
In order to test whether any LMXBs are formed in the 
field, we first adopted the simple hypothesis that:
\begin{equation}
N_{X} = \mu \frac{L^{TOT}_{GC}}{10^6L_\odot}  + \nu \left( \frac{L_V}{10^{10}L_{V\odot}} \right)
\label{eqn_nx_ngc}
\end{equation}
where $N_X$ is the total number of LMXBs (corrected for source detection
incompleteness), 
$L^{TOT}_{GC}$ is the total luminosity of the GC and $L_V$ is the V-band 
luminosity.
We performed this comparison only within the WFPC2 
FOV. For most of the galaxies in our sample, we tabulate in \gcpaper\ 
the specific luminosity of globular clusters,
defined as \Sl=100$L_{GC}/L_V$.
From 
Eq~\ref{eqn_nx_ngc}, $\mu$ and $\nu$ can be trivially obtained from a plot 
of \Sl\ versus the specific frequency of LMXBs, \Sxn, defined as
\Sxn=$8.55\times 10^{7} N_{X}/ (L_V/L_{V\odot})$ (assuming the absolute 
V-band magnitude of the Sun is 4.83: \citealt{maraston98a}).
This is similar to the method adopted by \citet{irwin05a}.

Adopting our preferred broken powerlaw XLF to compute $N_X$
(Table~\ref{tab_gcs2}), we show in 
Fig~\ref{fig_nx_ngc} \Sxn\ {\em versus} \Sl.
Taking account of errors on both x and y axes with a strategy
similar to that outlined in \citet{nr}, we fitted a straight line
through the data. Initially fixing $\nu$=0.0, \ie\ requiring that
all LMXBs form in GCs, we found a reasonable fit,
($\chi^2$/dof= 23.1/18) with $\mu=1.20\pm0.12$.
Allowing $\nu$ to fit freely, we found that the fit did not significantly
improve ($\chi^2$/dof=22.7/17) and we constrained $\mu=1.27^{+0.28}_{-0.22}$
and $\nu$ is $<1.8$.

For a typical galaxy with \Sl$\sim 0.2$, this implies that at most only $\sim$8\%
of the LMXBs could be formed in the field.
Since NGC\thin 5845 is something of an outlier, being the smallest galaxy
in our sample, we have investigated whether the results of our fit are
biased by including this galaxy. Excluding this galaxy, we obtained
similar results ($\chi^2$/dof=19.4/16, $\mu=1.18^{+0.28}_{-0.21}$ and
$\nu$ is $<3.4$), implying it does not bias the fit.
For comparison purposes, if we instead adopted a simple powerlaw
XLF, with $\beta=2.0$, we obtained a relatively poor fit for the case $\nu=0.0$
($\chi^2$/dof=39/18) with $\mu=2.80^{+0.32}_{-0.28}$.
Allowing $\nu$ to vary freely, the fit was similarly poor
($\chi^2$/dof=38/17) and we constrained $\mu=3.07^{+0.76}_{-0.56}$
and $\nu\le3.6$.

\subsection{GC LMXB retention fraction}
\begin{figure}
\centering
\includegraphics[scale=0.34]{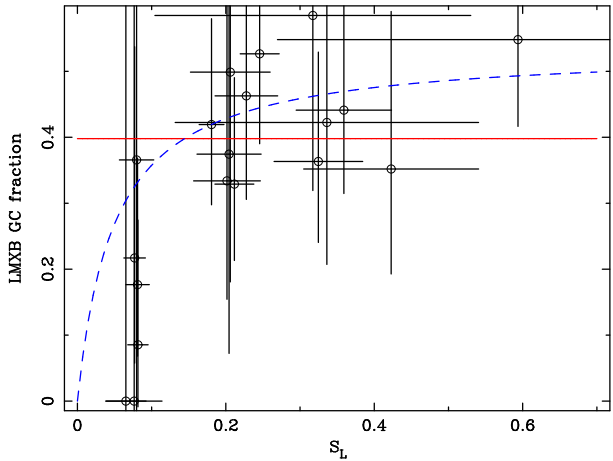}
\caption{Incompleteness-corrected fraction of LMXBs associated with
a GC, as a function of \Sl. Shown are the preferred relations for the case 
where all sources form in GCs (solid line; red) and where LMXB formation
in the field is proportional to the V-band luminosity (dashed line; blue).} 
\label{fig_pxgc}
\end{figure}
The probability that an X-ray binary born in a GC is retained by the GC, 
correcting for incompleteness, P(X,GC), is given by the relation:
\begin{equation}
N_{GC,X}-N_{false}=P(X,GC) \left(N_X-\nu \frac{L_V}{10^{10}L_{V\odot}}\right) f_i^X f_i^{GC}
\end{equation}
We estimated both P(X,GC) and $\nu$  by matching this model to the observed 
distribution of $N_{GC,X}$
{\em via} a fitting procedure similar to that described 
in \S\ref{sect_pgcx}.
Initially we tested the hypothesis $\nu=0.0$, \ie\ all LMXBs form in
GCs. We found that the data were well-fitted 
(\probhnought=81\%), and we constrained P(X,GC)=$0.40\pm0.06$.
Allowing $\nu$ to vary freely also gave a good fit (\probhnought=95\%)
with P(X,GC)=$0.53^{+0.11}_{-0.14}$ and $\nu=6.3^{+1.7}_{-5.4}$, in 
agreement with our results in \S~\ref{sect_field_lmxb}.
Fig~\ref{fig_pxgc} shows values of P(X,GC) computed directly from the
data (assuming all sources form in GCs), and the best-fit models.
For comparison, if we adopted values of $N_X$ and $f^X_i$ appropriate for a 
single powerlaw XLF (with $\beta=2.0$), we constrained 
P(X,GC)=$0.55^{+0.12}_{-0.14}$ and $\nu=15.1^{+3.9}_{-11.5}$.

\section{Discussion}
\subsection{The X-ray luminosity function}
\subsubsection{The shape of the XLF}
\begin{figure}
\centering
\includegraphics[scale=0.37]{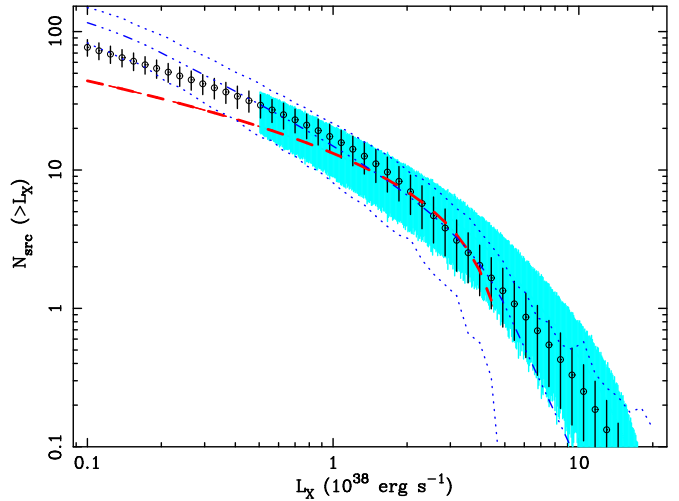}
\caption{A comparison of our preferred {\em cumulative} XLF model 
(circles; black. The error-bars denote the approximate 1-$\sigma$ uncertainty region)
with other XLFs from the literature. We have normalized the model to match
the observed number of sources with \lx$>L_{break}$ in the Milky Way.
We compare it with the cumulative XLF of Galactic LMXBs 
from \citet{grimm02} (dashed line; red). We have corrected this model for 
the difference in energy-bands used between these authors and our work,
under the assumption of a powerlaw spectrum with $\Gamma=2.0$. We also
compare to the ``universal'' LMXB XLF from
\citet[][blue; dash-dot-dot-dot line]{gilfanov04a}, with the blue dotted lines
delineating the approximate 1-$\sigma$ uncertainty region for the models.
We also show the best-fitting composite XLF for \citet[][light blue shaded region]{kim04b}.
The latter two models have been scaled arbitrarily for clarity.
We truncate both our model and that of \citeauthor{kim04b} above 
$2\times 10^{39}$ \ergps. Model curves are only shown corresponding to 
luminosity ranges over which there were data to constrain the model.
There is, in general, good agreement between the different models.
\label{fig_xlf_compare}}
\end{figure}
When corrected for source detection incompleteness, 
we found that the XLF of early-type galaxies is very uniform
in shape from object to object.
We found that a simple powerlaw fit to the XLF of each galaxy was 
formally acceptable, but the slope of the model correlated with
the source detection incompleteness, implying that the slope
flattens at low \lx.
Fitting the composite XLF we found that it is best modeled as a broken
powerlaw, with a break at $(2.21^{+0.65}_{-0.56})\times 10^{38}$\ergps, and 
negative differential logarithmic slopes 1.40$^{+0.10}_{-0.13}$ below the break 
and 2.84$^{+0.39}_{-0.30}$ above. Still, a single powerlaw model with 
slope, $\beta\simeq2.0$ was able to fit the XLF of each individual galaxy
formally about as well at this model.
A similar result was found by \citet{kim04b} for a smaller
sample of galaxies.

In Fig~\ref{fig_xlf_compare} we compare our best-fitting XLF, 
in cumulative form, with XLFs reported in the literature by a number of 
different authors. Considering the ranges for which each XLF has been 
found to be valid, we find excellent agreement between our best-fitting
functional form and those reported in the literature, even that reported
for Milky Way LMXBs by \citet{grimm02}. We note that the absolute normalization
of any of these models (except that of \citeauthor{grimm02}) is 
arbitrary. Despite this good agreement it is still possible that the XLF we measure 
is biased at low luminosities, where the data were dominated by only two galaxies,
NGC\thin 3115 and NGC\thin 4552, which may not be completely representative.
Unfortunately, this cannot be ruled out 
with our current data, but will require deep XLFs measured for  other
early-type galaxies to assess it. Nonetheless, most of our results
which depend on the XLF shape (for incompleteness correction)
do not appear to change qualitatively if we adopt an alternative XLF
shape which is much steeper at low luminosities.

There is some evidence in our results that there is real
scatter in the shapes of the XLF. 
In order to investigate this further, we considered the
very deep  XLFs determined in a few individual cases.
\citet{voss06a} report an XLF for Cen~A which shows a clear 
flattening below $5\times 10^{37}$ \ergps; the low-\lx\ $\beta$
is comparable to our measurements. Although it is difficult to
interpret a cumulative XLF due to the strong correlations between
data-points, it is interesting to note that NGC\thin 3379 shows
a significantly flatter XLF at low \lx\ than $\beta=2$
\citep{kim06a}, which would also be consistent with our
broken powerlaw model. Possible counter-examples, however, include M\thin 87,
in which the XLF can be fitted as a broken powerlaw but with
$\beta_1\sim 2$ \citep{jordan04a}. Fitting a differential (rather
than cumulative) XLF compiled from these authors' published source 
luminosities, we confirm this result, and find that our best fit to 
the composite XLF is marginally rejected. \citet{kim06a} find 
an XLF for NGC\thin 4278 which similarly appears to resemble 
an unbroken powerlaw, with $\beta\sim2$.
These results provide further evidence that 
there is genuine scatter in the shapes
of the XLFs between individual galaxies, 
in particular at low luminosities.
On average, though, we have shown that it must flatten 
at low \lx. Clearly further, deep studies of the LMXB
XLF in more galaxies are needed to investigate the origin of the scatter.

\subsubsection{The total luminosity of LMXBs} \label{sect_tot_lx}
\begin{figure}
\centering
\includegraphics[scale=0.37]{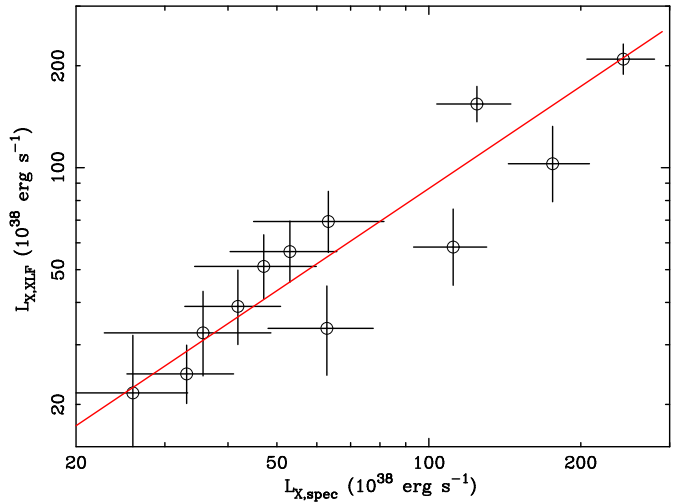}
\caption{A comparison between the X-ray luminosity of point-sources determined
from spectral-fitting (${\rm L_{X,spec}}$) and that determined from fitting 
the XLF (${\rm L_{X,XLF}}$) {\em within the same region} (see text). 
The best-fitting linear relation is shown. \label{fig_lx_compare}}
\end{figure}
\begin{figure*}
\centering
\includegraphics[scale=0.37]{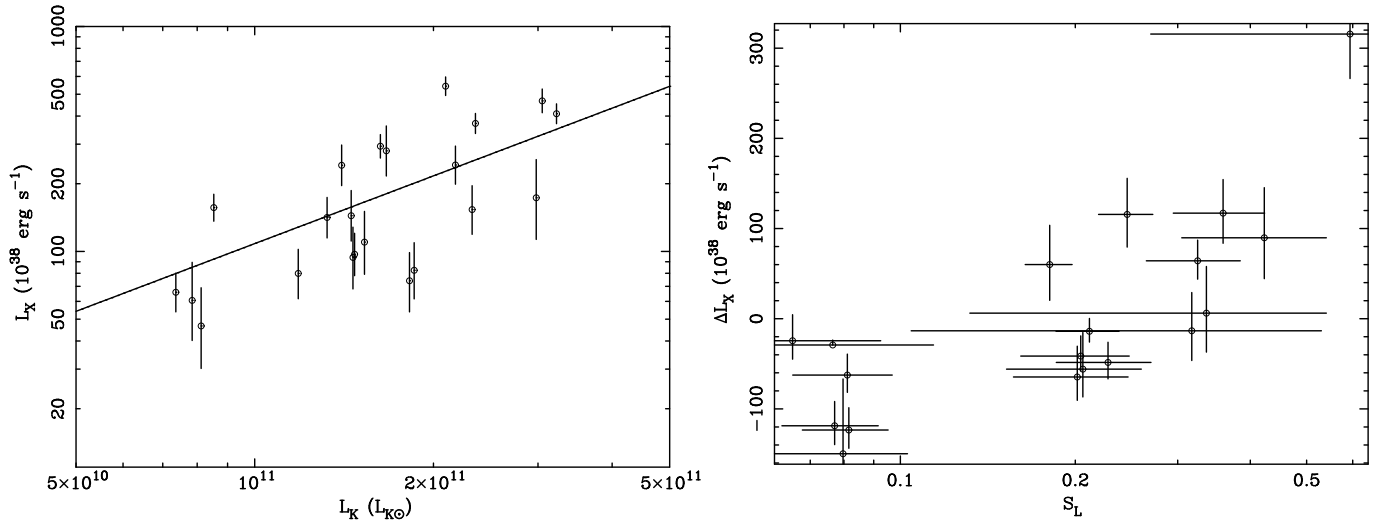}
\caption{{\em Left:} The integrated luminosity of the LMXBs in the galaxies,
shown as a function of K-band luminosity, adopting our preferred,
broken powerlaw XLF. We also show the mean straight line relation.
{\em Right:} The deviation of \lx\ from its the mean relation shown in the
left panel {\em versus} \Sl.} \label{fig_lx_v_lk}
\end{figure*}
In the present work, we chose to estimate the luminosity of the total LMXB
populations by extrapolation down to $10^{37}$ \ergps. This is, in part, 
justified by the excellent
agreement, at least to this luminosity, between our best-fitting composite 
XLF and that of LMXBs in the Milky Way (Fig~\ref{fig_xlf_compare}).
Another way to test the validity of this assumption is to adopt a procedure
similar to that outlined in \citet{irwin05a}. The ensemble average of 
LMXBs which are too faint to have been detected individually should contribute
a hard component to the spectrum of the diffuse emission in each galaxy. 
Therefore we  extracted, for each galaxy, a spectrum from the region in which
the XLF is computed (excluding regions around all detected point sources), 
and added the luminosity of the hard (unresolved LMXBs)
component to the the combined luminosity of the detected sources.
The details of the spectral extraction procedure and the treatment of the 
background are given in \citet{humphrey05a}. 

In practice, however, the procedure is rather complex for a number of 
reasons. First, some fraction of the point-sources are actually interloper
AGNs. All sources with apparent \lx $> 2\times 10^{39}$ \ergps\ were assumed to
be interlopers but, to minimize the effects of cosmic variance (which is 
dominated by the rare, bright point-sources), we disregarded these entirely. 
We estimated the expected {\em apparent} luminosity of interlopers over the 
flux range of interest from the model of \citet{tozzi01b}, and took into
account the fraction of the expected flux which had been accounted for
by the cosmic X-ray background component in our background modelling.
Since the LMXB emission arises from a finite number of discrete sources,
Poissonian statistics actually introduces an {\em intrinsic} uncertainty 
into the combined luminosity of these objects, irrespective of how 
accurately we can measure the total \lx. We estimated the magnitude of this
uncertainty {\em via} Monte-Carlo simulations, in which a fixed number of 
sources were randomly drawn from a luminosity distribution matching the 
measured XLF, and the total luminosity measured. For each set of simulations,
we were able to estimate the scatter (standard deviation) in the total 
luminosity  as a function of its expected value. For each galaxy, the scatter 
was added in quadrature with the measurement errors. 

Fig~\ref{fig_lx_compare} shows a comparison of the luminosity determined
from XLF fitting (${\rm L_{X,XLF}}$) and that determined from the 
above procedure (${\rm L_{X,spec}}$). In a significant fraction of the 
galaxies, we found that ${\rm L_{X,spec}}$ was extremely sensitive to the 
modelling or treatment of the background \citep[consistent with the assessment
of the systematic error budget in][]{humphrey05a}, and so these galaxies
have been omitted from the comparison. In our analysis, the background is 
determined through a modelling procedure, which we have determined to be 
more reliable than adopting the background ``templates''. We identified objects
in which the modelling is likely suspect, at least in the crucial high-energy
range, by examining the total flux of one of the modelled spectral 
components--- the powerlaw, which accounts for the unresolved fraction of the X-ray
background. Although this flux is subject to the effects of cosmic variance,
we should be able to resolve out a significant fraction of the brightest background 
sources (which contribute most to the cosmic variance). Therefore, we should reasonably 
expect that the flux of this modelled component should be {\em lower} than the 
total flux of the AGN contribution to the background found by, for example, 
\citet{tozzi01b},
even accounting for cosmic variance.
In half of the galaxies, we actually found that its flux was considerably higher
than expected from this argument, typically indicating systems in which our models
do not completely capture the shape of the high-energy background. In such circumstances,
${\rm L_{X,spec}}$ is almost certainly in error (even though the luminosity of the hot
gas emission, which has a much softer spectrum, is fairly well-determined), 
and so is omitted here.

We fitted a model of the form ${\rm L_{X,XLF} = \zeta L_{X,spec}}$ to 
these data. 
Since the XLF certainly extends to fainter LMXBs than $10^{37}$\ergps, we 
do not expect $\zeta$ to be exactly 1; its precise value depends on both
the shape of the XLF and the actual low-\lx\ cut for LMXBs.
Extrapolating our preferred XLF, for example, we expect $\zeta$ to be in the 
range 0.9--1.0. In contrast, for pure powerlaw XLF with $\beta=2.0$,
$\zeta$ may be as low as 0.70 if the XLF is unbroken down to $10^{36}$ \ergps.
Computing ${\rm L_{X,XLF}}$ for each galaxy with our best-fitting composite 
XLF, we obtained a good fit with this model ($\chi^2$/dof=11.1/11), with 
$\zeta=0.87\pm0.12$, which is consistent with expectations. 
Adopting ${\rm L_{X,XLF}}$ from the powerlaw fits to the 
XLF, we found $\zeta=1.06\pm0.16$, which implies that the XLF must be truncated
below a limit which is \gtsim $5\times 10^{36}$ \ergps.

Previous authors have reported
a strong correlation between the total luminosity of the X-ray
sources and the K-band luminosity of the galaxy \citep{gilfanov04a,kim04b}.
Using our larger sample of galaxies, we find a similar correlation
(Fig~\ref{fig_lx_v_lk}), corresponding to
\begin{equation} {\rm L_X^{LMXB}/L_K = 0.11\times 10^{30} erg s^{-1} L_{K\odot}^{-1}}
\end{equation} with a scatter of ${\rm 0.06\times 10^{30} erg s^{-1} L_{K\odot}^{-1}}$,
assuming our preferred XLF shape for each galaxy.
For comparison, if we assumed that the XLF is a powerlaw with $\beta=2.0$,
the mean and scatter (omitting IC\thin 4296, for which the error-bars
are large) became  ${\rm 0.17\times 10^{30} erg s^{-1} L_{K\odot}^{-1}}$
and ${\rm 0.09\times 10^{30} erg s^{-1} L_{K\odot}^{-1}}$, respectively,
in excellent agreement with \citet{kim04b}. In
both cases the statistical errors are considerably smaller than the 
intrinsic scatter. 

At face value this correlation
indicates that the integrated LMXB population is a good indicator of the 
stellar mass in the galaxy \citep{gilfanov04a}. However, as we show in
\S~\ref{sect_field_lmxb}, the majority, if not all, of the LMXBs must be formed in 
GCs. The relation between the 
numbers of LMXBs and the combined luminosity of the GCs does not show
 any scatter larger than the (albeit large) statistical errors. There is,
however, a correlation between luminosity and the numbers of GCs in a galaxy
suggesting that the correlation found in 
Fig~\ref{fig_lx_v_lk} may arise through this third parameter.
We show in the right panel of Fig~\ref{fig_lx_v_lk} the deviation of 
\lx\ from the mean relation with respect to \lk\ as a function of \Sl.
This shows a clear correlation, which would support this interpretation,
although this figure should be treated with caution since \Sl\ and \lx\ are 
not computed from self-consistent apertures.

\subsubsection{The high-\lx\ slope} \label{sect_slope}
\begin{figure*}
\centering
\includegraphics[scale=0.7]{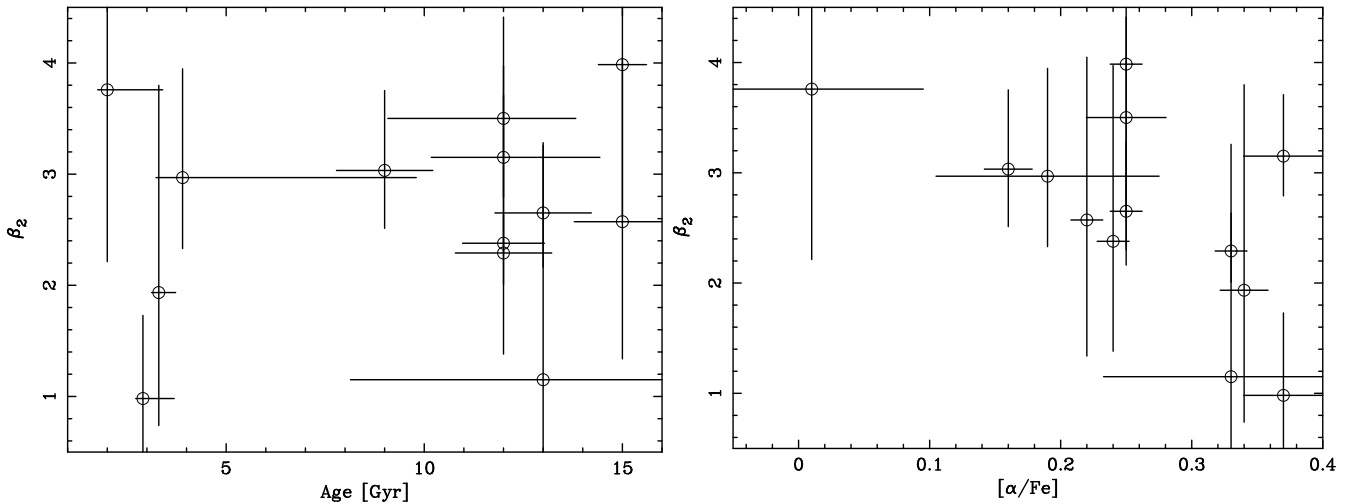}
\caption{The slope of the XLF at \lx$>2\times 10^{38}$\ergps\ versus age 
(left panel) and [$\alpha$]/Fe (right panel) of the stellar populations.} 
\label{fig_correlations}
\end{figure*}
Given the quality of the data, it was not possible to investigate how 
$L_{break}$ or ${\beta_1}$ vary between galaxies. 
However, we were able to investigate whether $\beta_2$, the high-\lx\
slope, does so, possibly providing further clues as to the origin of the
XLF shape.
Where it could be constrained, we allowed $\beta_2$ to fit freely and the 
results are shown in Table~\ref{tab_xlf}.
We searched for correlations between it and various interesting
parameters such as age, metallicity and [$\alpha$/Fe] for the stellar population,
\Sl\ and the peak colour of the GC distribution. 
The ages and metallicities listed in Table~\ref{table_obs} 
were computed as described in \citet{humphrey05a} and \citet{humphrey06a}.
For those systems for which the age had been fixed to 12~Gyr in our previous
analysis we relaxed this constraint when estimating the age.

If the high-luminosity sources are dominated by short-lived objects
forming in the field \citep[\eg][]{wu01a}, we would expect to see
a correlation between $\beta_2$ and the stellar 
age. \citet{ivanova06b} constructed a model for the high-\lx\
XLF slope, which they related to the transient duty cycle. In this picture,
if mass-transfer is dominated by red giant or white dwarf donors,
$\beta_2$ should also depend on age.
However, we did not find any evidence of a 
correlation (prob(no correlation)=70\% for Kendal's $\tau$-test,
and we obtained similar results with Spearman's rank test) between
$\beta_2$ and age (Fig~\ref{fig_correlations}).

In contrast we did find a marginally significant 
correlation between [$\alpha$/Fe] and $\beta_2$, as shown in Fig~\ref{fig_correlations}, 
for which prob(no correlation)=3--8\% for the different tests. 
To investigate whether 
such a correlation is robust in the presence of the large statistical errors,
we investigated the effect of randomly scattering the data-points within their
error-bars. We found that 56\% of the resulting simulations gave a significant
correlation (in comparison to the 5\% expected if the correlation is not robust).
Given the number of different trials we performed, the correlation between
[$\alpha$/Fe] and $\beta_2$ should be considered tentative at best, and 
needs to be confirmed with a larger data-set. Nonetheless, 
taking it at face value, it is intriguing.
One possible cause of such a correlation is a systematically-varying
IMF. A more top-heavy IMF would increase [$\alpha$/Fe]  
and also affect the 
compact object demographics. If there are more massive objects,
it is unsurprising if more sources are observed which are able to sustain
high \lx. 
Alternatively, enhanced [$\alpha$/Fe] in the stars may significantly
affect how mass transfer operates, increasing the population
of X-ray bright LMXBs. However, neither the peak GC colour nor
overall metallicity appear to correlate with $\beta_2$, and it is not clear
why [$\alpha$/Fe] should have a stronger impact on the accretion process
than total metallicity.
Clearly both of these scenarios require that either the high-\lx\ sources
are systematically biased to form in the field (since most LMXBs do not), 
or the variations in stellar [$\alpha$/Fe] 
(and/ or the IMF) must also be reflected 
in the GC populations.

\subsubsection{The meaning of the break}
Early results suggesting the presence of an XLF break at $\sim 2\times 10^{38}$ \ergps\
have been called into question due to the failure of the authors
to correct properly for source detection incompleteness \citep{kim04b}. 
Although we now confirm the presence of such a break even when the 
data have been corrected for incompleteness,
{\em a break was not required statistically in any individual galaxy}.
The significance of the feature in some early analyses may have been
exaggerated, therefore, by incompleteness effects.

It has been suggested \citep{sarazin01} that the break
represents a division between neutron-star and black-hole binary
systems, with objects above this limit being entirely black hole binaries,
and those below the limit a mixture of black hole binaries and neutron-stars. 
It is difficult to reconcile this hypothesis, however, with the shape of the
XLF. Most systems are not observed 
close to their Eddington limit; for example all neutron-star LMXBs
do not have \lx$\sim 2\times 10^{38}$\ergps. Therefore, the black-hole
binary XLF would be expected to extend below this limit and,
crucially, would not be expected to exhibit any feature at 
this characteristic luminosity. An extrapolation of the 
high-\lx\ slope
(comprising only the putative black hole population) dramatically
over-estimates the numbers of sources at low \lx, since the 
XLF {\em flattens} below the break, effectively ruling out this picture.
Although \citet{sarazin01} actually recognized this problem, they argued
there must be different source populations either side of the break. However,
this would still, in fact, require the black hole binary XLF to ``care'' about the 
Eddington limit of a neutron-star, which is not expected.
\citet{bildsten04a} explained both the break and low-\lx\ slope of the XLF
found by \citet{kim04b} in terms of ultra-compact X-ray binaries. However,
the break luminosity we detect is significantly lower and, more 
problematically, the low-\lx\ slope predicted ($\beta \sim 1.8$) is 
much steeper than we detect.

It is worth noting that a further problem with interpreting the break as 
the Eddington limit is, of course, that the point source \lx\ we measure
in the \chandra\ band is certainly an underestimate of the bolometric
luminosity. For example, for the representative Galactic neutron-star 
X-ray binaries 
X\thin 1916-053 and X\thin 1624-490, an extrapolation of the best-fitting
models for broad-band BeppoSAX observations \citep{church98a,balucinska00b}
implies the 0.3--7.0~keV flux underestimates that in the 0.1--200~keV band by 
factors $\sim$2.8 and 1.5, respectively, and obviously the bolometric flux
is underestimated by even larger factors.

Another possible explanation for the break is that it 
arises from an evolving LMXBs population born in a putative recent
burst of star formation  \citep{jeltema03,wu01a}. The absence of 
any obvious correlation between the high-\lx\ shape and the age
of the stars seems to rule out this model. 
Furthermore, given the short lifetime of an X-ray binary (\ltsim $10^8$yr)
this model seems unlikely to be able to sustain a break at 
$2\times 10^{38}$\ergps\ in a typical galaxy with a $\sim$12~Gyr
stellar population.  In any case, we did not find
any convincing evidence of a break which systematically evolves
from galaxy-to-galaxy
and most, if not all, LMXBs are formed in GCs.

The well-studied population of Milky Way LMXBs provides some intriguing clues 
as to what may be the origin of the break. 
Galactic LMXBs are known to be highly variable \citep[\eg][]{liu01,grimm02}.
In particular, all confirmed LMXB black hole binaries in the Milky Way 
are transient sources \citep{mcclintock03}, and there is evidence
that a significant portion of LMXBs in early-type galaxies may also
be transient \citep[\eg][]{kraft01}.  
Different classes of LMXB, which exhibit strikingly
different patterns of variability occupy distinct luminosity bands
\citep[\eg][]{hasinger89}, so that the XLF is not expected to be 
``scale-free''.
It is striking that the so-called ``Z-track'' sources, which comprise
the majority of the most luminous persistent LMXBs in the Milky Way
have \lx $\sim$1--3$\times 10^{38}$\ergps\ \citep{grimm02}, measured in 
the \chandra\ band, in good agreement 
with $L_{break}$ in the early-type galaxy XLF.
If, as in the Milky Way, sources brighter than this limit are predominantly
lower-\lx\ sources exhibiting brief periods of outburst or flaring, this would
naturally cause a break. This
 can be explicitly tested by looking for characteristic patterns of 
(spectral) variability in the brightest LMXBs. 
In the current work there were, unfortunately, insufficient photons to 
allow us to do this.
One possible counter-argument to this picture is the observation by
\citet{irwin06a} that no transient behaviour was seen in the 
brightest sources in NGC\thin 1399 and M\thin 87, over a timescale of 
$\sim$a few years, effectively placing a limit on transient behaviour in
these objects of $\sim$50~yr. Although these objects are obviously 
not short-term transients, in no way does this work suggest they
are {\em not} transient over periods of a few decades or longer, 
and so our interpretation still holds. In any case, such long-term 
transients appear very rare (as seen in the Milky Way).

As a corollary, it is interesting to note that source variability may
have other impacts on the shape of the XLF. The 
characteristic patterns of short term variability \citep[\eg][]{white95} 
are superimposed on long term secular evolution in the mass-transfer rate 
\citep[\eg][]{podsiadlowski02}, so that the locus of an individual
source in the XLF should evolve with time. It is possible, therefore, that the 
time-averaged lightcurve may be the dominant factor in determining the 
shape of the XLF, which might explain the comparative uniformity in its
shape from galaxy to galaxy. Obviously compact object demographics 
must play a role since black-hole binary systems exhibit, on average,
slightly different properties to neutron-star systems \citep{tanaka95}.
Therefore differences in the IMF might still produce 
the possible correlation we see between [$\alpha$/Fe] and $\beta_2$.

\subsection{The GC-LMXB connection}
We find a strong correlation between the number of GCs in the galaxy
and the number of LMXBs, placing tight constraints on the fraction
of sources formed in the field. We find that the number of LMXBs formed
in the field is $<1.8 \times 10^{-10} L_V/L_{V\odot}$, implying
\ltsim 10\% of the LMXBs in a galaxy with \Sl $\sim 0.2$.
In fact, the data are consistent with the hypothesis that {\em all}
the LMXBs are formed within GCs. Ideally we would have performed
this comparison using the K-band, rather than the V-band light, 
since it better traces  the mass of the underlying stellar population.
However, since our best-fitting relation passes through the origin,
this indicates the number of LMXBs is consistent with being 
exactly proportional to the combined V-band luminosity of the GCs.
Since the GCs are expected to be uniformly old stellar systems 
(\eg.\ \gcpaper), the V-band mass-to-light ratio should not 
vary considerably between the galaxies. Scaling both the number of 
LMXBs and the luminosity of GCs for each galaxy 
by the same arbitrary factor (\eg\ the 
K-band luminosity of the galaxy; hence computing \Sxn\ and \Sl\
using the K-band, rather than V-band luminosity) should not affect 
the quality of a ``y$\propto$x'' fit to the data. Although the exact
numerical relations we derive depend on the shape of the XLF we adopt
for incompleteness correction, qualitatively our conclusions are 
found to be insensitive to the exact XLF shape.

Our conclusions are significantly different from those of \citet{irwin05a},
who, using a smaller sample of galaxies, argued that the total
{\em luminosity} of the LMXBs divided by the optical luminosity
of the galaxy is proportional to \Sn\ {\em plus a constant offset}. 
We suspect the reason for this discrepancy arises from the way in which
the luminosity of undetected LMXBs was estimated.
\citet{irwin05a} computed the luminosity of unresolved LMXBs
from the high-energy (2.0-6.0~keV) portion of the diffuse emission 
spectrum in each galaxy (excepting NGC\thin 1399). This implicitly 
assumes that there is no contribution from the hot ISM at these energies,
which for a number of the galaxies in the sample may not be strictly
true \citep[\eg][]{humphrey05a,humphrey04b}. 
More problematically, the determination
of the unresolved source flux from the spectrum is very sensitive to the 
treatment of the  background (\S~\ref{sect_tot_lx}). Although 
\citeauthor{irwin05a} took a local background from the outer parts of the 
S3 chip, this fundamentally requires that there is no extended diffuse emission
in any of the galaxies, which is not true for all of the galaxies in the 
sample \citep[\eg][]{humphrey04b}, and requires careful 
treatment of vignetting effects. In fact, he found that, when attempting
spectral fitting, the models systematically underestimated the high-energy
spectrum, in contrast to our results when carefully modeling
the background \citep{humphrey05a}, suggesting that the background may
not have been completely accounted for.

We found that $\sim$1 currently active (\lx$> 10^{37}$\ergps)
LMXB was formed per $10^6$\lsun\ V-band integrated
luminosity of GCs, of which $\sim$40--50\% of the LMXBs are retained 
within the GC. Approximately 7\% of the GCs within the WFPC2 field
contain an active LMXB, in rough agreement
with the fraction seen in the Milky Way and, possibly, M\thin 31 
\citep{distefano02a}. 
If the lifetime of an LMXB is $\tau_{L}$ and the fraction of the 
time it is active with \lx$> 10^{37}$\ergps\ is $F_{d}$, our 
results imply an LMXB formation rate in GCs of 
${\rm \sim 10^{-14} L_{V\odot}^{-1} yr^{-1} (\tau_{L}/10^8 yr)^{-1} F_{d}^{-1}}$, or ${\rm \sim 1.5 Gyr^{-1} (\tau_{L}/10^8 yr)^{-1} F_{d}^{-1}}$ per GC,
averaging over the mean GCLF.

Another piece of evidence which supports the hypothesis that all LMXBs
form in GCs is the
good agreement between the spatial distribution of the LMXBs and the GCs,
when one considers an appropriately-weighted combination of the red and blue
GCs.
We find no evidence of a difference in the spatial distributions of LMXBs
found in GCs and the field implying that field LMXBs are most likely 
ejected only with modest kick velocities from GCs. 
With a smaller sample of galaxies \citet{kim05a} found similarities in the
distributions of the LMXBs in the field and in GCs, consistent with our
result, but found that the LMXB distribution was far steeper than the 
GCs. These authors'
optical data covered a larger radial range than ours, and so it is possible
the agreement we find with our data is an artifact of our smaller field of view.
However, we note that these authors did not correct for the spatial variation
in source detection incompleteness, which affects radial profiles of both
the LMXBs and the GCs, and may have played a role in this discrepancy.
In contrast \citet{kundu07a} report, based on a sample of  5 galaxies, 
that the LMXB-hosting GCs are distributed similarly to the GCs, whereas
field LMXBs have a more centrally-peaked distribution.
To some extent, we suspect that small number statistics also plays a role 
in the discrepancies between these authors' work and ours.
The sample of \citet{kundu07a}, for example, contains the S0 galaxy 
NGC\thin 3115, which had the most centrally-peaked LMXB distribution in our 
sample.
Furthermore, these authors compare the distribution of LMXBs and that of 
{\em all} the GCs, which is more extended than an
appropriately-weighted combination of red and blue GCs
(see \S~\ref{sect_gc_dist}).

Having corrected for source detection incompleteness in both the 
LMXB and GC populations, we found little evidence that the fraction of 
LMXBs coincident with GCs vary from galaxy to galaxy. This is in sharp
contrast to past studies \citep{juett05a}, which argued for 
a correlation between this fraction and the specific frequency of 
globular clusters, which implies a significant fraction of the LMXBs
are formed in the field. This study, however, drew upon literature 
results which did not always treat the data consistently. 
In particular, literature values of \Sn\ were used, which were
typically not computed in the same aperture as was used for LMXB-GC 
matching. Since the distribution of GCs as a whole is flatter than
the optical light (or the distribution of LMXBs, which follow the 
more centrally-peaked red GC distribution), it is crucial that \Sn\ 
is computed within this aperture. Adopting locally-computed
specific frequencies the correlation of
\citeauthor{juett05a}
is seriously degraded, as pointed out by \citet{kim05a}.
 Furthermore, when one corrects for 
incompleteness, there is little evidence of any statistically significant
variation from galaxy-to-galaxy.

We found that the probability a given GC contains an 
LMXB is proportional to its luminosity to the power of $1.01\pm0.19$
and its metallicity to the power of $0.33\pm0.11$.
These values are very similar to recent results obtained by
\citet{sivakoff06a}, \citet[][for M\thin 87]{jordan04a} 
and the sample of  \citet{smits06a}.
In order to interpret these results in terms of 
physical models, it is necessary to recast  Eqn~\ref{eqn_pgcx} 
in terms of $M_{GC}$, the mass of the GC. For the range of stellar
metallicities implied by the colours of LMXB-hosting GCs
($[Fe/H]\sim$-1.6--0.25), the stellar M/\lv\ ratio actually depends
on metallicity. To illustrate this point, we adopted  
M/\lv\ ratios for an assumed 10~Gyr population
with a Kroupa IMF from \citet{maraston98a}\footnote{Using the
updated model-grids made available by the author at http://www-astro.physics.ox.ac.uk/$\sim$maraston/ Claudia's\_Stellar\_Population\_Models.html}
and  found $M/L_V \sim 3.5 Z_{Fe}^{0.19}$, which is accurate to better than 
$\sim$10\% for the range of interest. Inserting
this into Eqn~\ref{eqn_pgcx}, we obtained 
$P(GC,X)\simeq 0.023 (M_{GC}/M_0)^{1.01\pm0.19} (Z_{GC}/Z_0)^{0.14\pm0.12}$, 
where $M_0= 1.6\times 10^5$\msun. 

\citet{maccarone04a} proposed an irradiation-induced
wind model to account for an excess of red GCs harbouring LMXBs,
which predicts an approximately powerlaw dependence on
metallicity, with an exponent $\sim$0.39. This is stronger
than the dependence we obtained (differing at $\sim 2\sigma$). 
Furthermore, we 
note that the predicted enhanced absorption in the blue-GC LMXBs than
red-GC LMXBs is not observed in our data.
It is interesting to note that \citet{maccarone04a} also estimated that
the effects of metallicity on stellar radius should lead to a dependence
$P(GC,X) \propto Z_{GC}^{0.12}$, which is in good agreement with our 
estimate above.
We stress that our value of the metallicity exponent is 
derived assuming that all of the colour variation between GCs arises
solely from metallicity effects, ignoring the effects of age, which
would affect both the M/L ratio and the conversion from GC colour to
metallicity. Clearly further multi-colour or deep spectroscopic
observations to break the age-metallicity degeneracy are needed to 
investigate this further.

\section{Summary and conclusions}
Using a sample of 24 galaxies observed with \chandra\ we found, in summary:
\begin{enumerate}
\item Correcting for source detection incompleteness, the point-source 
XLF of  individual galaxies is consistent with a single powerlaw,
with $\beta \sim$2.0. We find that $\beta$ correlates with incompleteness,
indicating that the XLF steepens at low \lx.
\item The composite XLF of all the galaxies is best-fitted by a 
broken powerlaw with a break at $2.21^{+0.65}_{-0.56}\times 10^{38}$\ergps\
and slopes $1.40^{+0.10}_{-0.13}$ and $2.84^{+0.39}_{-0.30}$ above and below
the break, respectively. 
\item The shape of the XLF is inconsistent with the break
representing a strict division between black hole and neutron-star systems.
However, in common with the Milky Way, it may represent the luminosity of the 
brightest persistent LMXB, so that sources with higher \lx\ are 
exhibiting flaring or in transient outburst, naturally producing a break.
\item The slope of the XLF at high luminosities does not correlate with the 
stellar population age, but shows evidence of a weak correlation with
[$\alpha$/Fe], suggesting a possible effect of the IMF on the XLF.
\item We found that the combined LMXB luminosity is approximately proportional
to \lk\ of the galaxy. However, there is significant intrinsic scatter in the 
relation, such that 
${\rm L_X^{LMXB}/L_K}$ has a mean and standard deviation of 
${\rm 0.11}$ and 
${\rm 0.06\times 10^{30} erg s^{-1} L_{K\odot}^{-1}}$.
\item We find no difference in the XLF or composite spectra between LMXBs in GCs and
 those in the field.
LMXBs in red and blue GCs have essentially the same spectrum, ruling out
the irradiation induced model of \citet{maccarone04a}.
\item Correcting for incompleteness, GC colour and GCLF effects, we find that 
the probability a GC hosts an LMXB with \lx $>10^{37}$ \ergps\ is 
$\sim 6.5$\%, and the probability an LMXB is in a GC is $\sim$40\%. 
These do not vary from galaxy to galaxy. 
\item  The probability that a GC hosts an LMXB depends on luminosity and 
metallicity to the powers $\alpha=1.01\pm0.19$ and $\gamma=0.33\pm0.11$,
respectively. When the metallicity
dependence of the stellar M/L ratio is taken into account, the tendency for
red GCs preferentially to host LMXBs may be consistent solely with
metallicity influences on stellar radii.
\item The specific frequency of LMXBs is  proportional to the \Sl\ without
a significant constant offset,
implying that all LMXBs form in GCs. Our results imply an LMXB formation rate
of  ${\rm \sim 1.5 Gyr^{-1} (\tau_{L}/10^8 yr)^{-1} F_{d}^{-1}}$ per GC
(where $\tau_L$ is the LMXB lifetime and $F_d$ the fraction of its life spent
with \lx $>10^{37}$ \ergps), and put an upper limit of 
$\sim1.8\times 10^{-10} L_V/L_{V\odot}$ LMXB with \lx $> 10^{37}$ \ergps\
formed in the field.
\item This is further supported by the fact that the LMXB distribution 
closely follows the  distribution of red GCs, when incompleteness is 
taken into account. Furthermore, the spatial distributions of GC and field 
LMXBs are essentially the same.
\item With the present data we cannot completely rule out that the two galaxies
which dominate the XLF at the lowest \lx\ are not representative, which could
distort the XLF shape in that region. Although this would have an impact on
our numerical results, qualitatively our conclusions appear insensitive to the 
low-\lx\ slope.
\end{enumerate}

\acknowledgements
We would like to thank Fabio Gastaldello and Luca Zappacosta for interesting
discussions. We would also like to thank Jimmy Irwin for providing comments
on the draft, and Rasmus Voss for discussions.
This research has made use of data obtained from the High Energy Astrophysics 
Science Archive Research Center (HEASARC), provided by NASA's Goddard Space 
Flight Center. In addition, we made use of data obtained from the MAST
archive. This research has also made use of the 
NASA/IPAC Extragalactic Database (\ned)
which is operated by the Jet Propulsion Laboratory, California Institute of
Technology, under contract with NASA, and the HyperLEDA database
(http://leda.univ-lyon1.fr).
Support for this work was provided by NASA under grant 
NNG04GE76G issued through the Office of Space Sciences Long-Term
Space Astrophysics Program.

\appendix
\section{Source lists}
We include in Table~\ref{tab_sources} the point-source lists of all 
the galaxies in the sample.
We include only those sources within \dtwentyfive\ of each galaxy,
which should mitigate against contamination by background AGN.
For each source, we provide the source name, which incorporates the 
coordinate information, the total number of counts detected, the 
0.3--7.0~keV X-ray luminosity (\lx), the two hardness ratios HR1 and HR2
(\S~\ref{sect_hardness}) and the projected distance from the galaxy centre
($\Delta_R$). For those LMXBs within the appropriate WFPC2 field, we 
list under ``GC'' the GC catalogue number given in \gcpaper\ for any which 
match GCs. If the source has no match, we report ``None'', and if the source
is not in the WFPC2 field, it is marked by an ellipsis.

\bibliographystyle{apj_hyper}
\bibliography{ms}

\setlength{\hoffset}{-15mm}
\renewcommand{\tabcolsep}{2.0mm}
\begin{deluxetable}{lllllll}
\tablewidth{0pt}
\tablecaption{LMXB Source lists \label{tab_sources}}
\tablehead{
\colhead{Source }&\colhead{Counts }&\colhead{\lx}&\colhead{HR1}&\colhead{HR2}&\colhead{$\Delta_R$}&\colhead{GC} \\
\colhead{ }&\colhead{ }&\colhead{($10^{38}$\ergps)}&\colhead{ }&\colhead{ }&\colhead{(\arcmin)}&\colhead{ }}
\startdata
\multicolumn{7}{c}{\bf IC 4296} \\ \hline
CXOU J133638.2-335755 &$27.$ &$26.\pm 12.$ &$0.43\pm 0.26$ &$0.66\pm 0.49$ &$0.17$ &\ldots \\
CXOU J133639.4-335713 &$25.$ &$24.\pm 10.$ &$0.16\pm 0.27$ &$0.68\pm 0.40$ &$0.74$ &\ldots \\
CXOU J133639.7-335751 &$17.$ &$19.\pm 12.$ &$0.31\pm 0.60$ &$-0.30\pm 0.53$ &$0.18$ &\ldots \\
\ldots & \ldots & \ldots & \ldots & \ldots & \ldots &\ldots 
\enddata
\tablecomments{Catalogue of detected sources in each galaxy (see text). The full version of this table is 
available in the Electronic version of the paper.}
\end{deluxetable}

\setlength{\hoffset}{0mm}

\end{document}